\documentclass[journal,twoside]{IEEEtran}

\usepackage{amsmath}
\usepackage{cite}
\usepackage{epsf}
\usepackage{subfigure}

\newtheorem{lemma}{Lemma}

\def\vec#1{{\bf #1}}

\def\Tr#1{\mbox{Tr}\left\{ #1 \right\} }

\def\bm#1{{\mbox{\boldmath{$#1$}}}}

\begin{document}
\bstctlcite{BSTcontrol}

\title{Capacity of Differential versus Non-Differential Unitary Space-Time Modulation for MIMO channels} 

\author{Aris L. Moustakas,~\IEEEmembership{Senior Member,~IEEE,} %
Steven H. Simon and Thomas L. Marzetta,~\IEEEmembership{Fellow,~IEEE}%
\thanks{A. L. Moustakas (email: arislm@phys.uoa.gr), S. H. Simon (email:
shsimon@bell-labs.com) and T. L. Marzetta (email:
tlm@research.bell-labs.com) are with Lucent Technologies.}}
\date{September 24, 2005}



\maketitle

\begin{abstract}
Differential Unitary Space-Time Modulation (DUSTM) and its earlier
nondifferential counterpart, USTM, permit high-throughput MIMO
communication entirely without the possession of channel state
information (CSI) by either the transmitter or the receiver.  For
an isotropically random unitary input we obtain the exact
closed-form expression for the probability density of the DUSTM
received signal, which permits the straightforward Monte Carlo
evaluation of its mutual information.  We compare the performance
of DUSTM and USTM through both numerical computations of mutual
information and through the analysis of low- and high-SNR
asymptotic expressions.  In our comparisons the symbol durations
of the equivalent unitary space-time signals are both equal to
$T$, as are the number of receive antennas $N$. For DUSTM the
number of transmit antennas is constrained by the scheme to be $M
= T/2$, while USTM has no such constraint.  If DUSTM and USTM
utilize the \emph{same number of transmit antennas} at high SNR's
the normalized mutual information of the differential and the
nondifferential schemes expressed in bits/sec/Hz are
asymptotically equal, with the differential scheme performing
somewhat better, while at low SNR's the normalized mutual
information of DUSTM is asymptotically twice the normalized mutual
information of USTM. If, instead, USTM utilizes the optimum number
of transmit antennas then USTM can outperform DUSTM at
sufficiently low SNR's.
\end{abstract}


\section{Introduction}
\label{Introduction}

\PARstart{C}{onsiderable} volume of work has followed the
prediction\cite{Foschini1998_BLAST1, Telatar1995_BLAST1} that the
use of multiple antennas in transmitting and receiving signals can
result to substantial increases in information throughput. The
underlying assumptions of this effort have been that the receiver
knows the channel through some training scheme and that the
channel coefficients are statistically independent. In this case
and for large signal to noise ratio $\rho$, the capacity is
roughly
\begin{equation}\label{eq:cap_large_rho_coherent}
C_{coh} \approx \min(M,N)\log_2\rho \,\, \mbox{bits/sec/Hz}
\end{equation}
where $M$, $N$ are the numbers of transmitting and receiving
antennas.

In a typical mobile wireless communication system the channel
coefficients vary continuously, following a Jakes-like
distribution. Thus one can only assume that the channel is
approximately constant over only limited periods of time. Hence,
especially for large transmitting antenna numbers, training will
require a substantial fraction of the coherence time of the
channel and thus hamper the data throughput rates. To address this
problem, Marzetta and Hochwald\cite{Marzetta1999_USTM,
Hochwald2000_USTM} investigated the scenario where the receiver
has no a-priori channel knowledge. In addition to the conventional
additive Gaussian noise, this channel has also multiplicative
noise, corresponding to the channel matrix, which is also assumed
to be Gaussian. This is a
 ``non-coherent'' channel, as opposed to the additive white Gaussian noise channel with known
(and static) ``coherent'' channel coefficients at the receiver. In
an elegant group-theoretic approach, Zheng and Tse
\cite{Zheng2002_CommunicationOnGrassmannManifold} found the
capacity of this channel to scale as
\begin{equation}\label{eq:cap_large_rho_incoherent}
C_{incoh} \approx M^*(1-M^*/T)\log_2\rho \,\, \mbox{bits/sec/Hz}
\end{equation}
for large $\rho$, where $M^* = \min(M,N,T/2)$ and $T$ is the
number of time intervals over which the channel is static. A
similar approach was developed independently by
\cite{Sengupta2000_BLASTWorkshop}. This implies that for fixed
$T$, there is no need to use more than $M=T/2$ transmitters.

To take advantage of the constancy of the channel over $T$ time
intervals, \cite{Hochwald2000_USTM} proposed to encode the signal
using $T\times M$ isotropic unitary matrices. In this encoding,
called isotropic unitary space-time modulation (USTM), a symbol
can be spread not only over $M$ antennas, but also over $T$ time
intervals. Some analytic results on the mutual information of USTM
already exist. In particular, it has been shown that for $T\gg M$
\cite{Marzetta1999_USTM} and for $M<\min(N,T/2)$ and large $\rho$
\cite{Zheng2002_CommunicationOnGrassmannManifold} the optimal
input distribution is isotropic random unitary, i.e. that of USTM.
Thus the asymptotic capacity is equal to the mutual information,
as in \cite{Marzetta1999_USTM,
Zheng2002_CommunicationOnGrassmannManifold, Telatar1995_BLAST1}.
Recently, Hassibi and Marzetta \cite{Hassibi2002_USTM}
analytically calculated  the received signal distribution and thus
were able to numerically evaluate the mutual information of USTM
for a variety of $M$, $T$, $N$ and $\rho$, confirming some of the
above asymptotic results. More recently,
\cite{Simon2004_CapacityOfCorrRandomWishartMatrices} generalized
the received signal distribution to channels with spatial
correlation.

In the case of USTM it is implicitly assumed that, after $T$
symbols the channel completely changes. In contrast, differential
phase-shift keying (DPSK) \cite{Proakis_book} has been used
extensively to take advantage of the continuous slow-varying
nature of the channel, without needing to perform any training. In
this scheme, each transmitted symbol is encoded into a
phase-difference from the previous symbol.

In \cite{Hochwald2000_DUSTM, Hughes2000_DUSTM}, the concept of
differential modulation was extended to multi-antenna systems. In
this method, called differential unitary space-time modulation
(DUSTM), the signal is encoded over $M$ transmitting antennas and
$M$ time intervals using an $M\times M$ unitary matrix. In each
successive $M$ time intervals, the transmitter encodes the input
signal by multiplying a $M\times M$ unitary matrix to the unitary
matrix transmitted during the previous $M$ time intervals over the
$M$ antennas and then transmits the matrix product. In turn, the
receiver decodes the signal by comparing the received signal from
the $M$ antennas and $M$ time intervals to that received over the
previous $M$ time intervals. Thus this scheme requires no training
and assumes that the channel is fixed over $T=2M$ time intervals.
The technique of DUSTM can be applied to the mathematically
identical space-frequency channel that appears during a single
OFDM symbol interval, resulting in a variation called differential
unitary space-frequency modulation (DUSFM)
\cite{Marzetta2003_DUSFM}.

Despite its importance in practical applications
\cite{Hochwald2000_DUSTM}, no analytic results are available
regarding the mutual information of DUSTM and its comparison with
USTM for $T=2M$. The main obstacle has been the difficulty in
integrating over exponentials of unitary matrices. This is a
problem that was tackled in the 80's by high-energy physicists in
analyzing the nuclear strong interactions (quantum
chromodynamics). Due to the $SU(3)$ symmetry of these interactions
their fluctuations can in certain cases be represented by unitary
matrices. Thus to integrate them out, one needs to make use of
such integrals of exponentials of unitary matrices. In this paper
we apply these results derived by \cite{Brower1981_LatticeQCD} to
the context of DUSTM. The methodology of the proof in
\cite{Brower1981_LatticeQCD} is based on mapping the original
problem to a diffusion problem of eigenvalues, which has a
differential equation that can be solved. Given its complexity it
will not be discussed at all in this paper. However, the
interested reader is referred to
\cite{Simon2004_CapacityOfCorrRandomWishartMatrices}, where some
of us apply the method of character expansion to derive the same
result and apply it to the capacity of Ricean MIMO channels. In
the present paper, we get the following results:
\begin{enumerate}
\item We analytically calculate the received signal distribution
for the case of DUSTM (see section \ref{sec:Closed-form solution
pX_DUSTM}).
\item Using this received signal distribution, we evaluate
numerically $I_{DUSTM}$, the mutual information of DUSTM for a
variety of $M$, $N$ and $\rho$, and compare it to $I_{USTM}$, the
mutual information of USTM setting $T=2M$. At low $\rho$ we find
that the two mutual informations for the same $M$, $N$, $T=2M$ are
nearly identical. This implies that the number of bits {\em per
symbol} i.e. $I_{DUSTM}/M$ is twice $I_{USTM}/T = I_{USTM}/2M$. In
contrast, at large $\rho$ the number of bits/symbol of the two
schemes approach each other, but with $I_{DUSTM}/M>I_{USTM}/T$.
\item We compare the maximum with respect to $M$ of the two mutual informations per
symbol. For fixed $M$, $N$, $T=2M$, we find that while at large
$\rho$ we have $\max_{M^*\leq
M}I_{DUSTM}(M^*,N,\rho)/{M^*}>\max_{M^*\leq
M}I_{USTM}(M^*,N,\rho,T)/T$, at small $\rho$ the opposite
inequality holds.
\item We back the above numerical results by providing
 expansions of the mutual information for both small and large
$\rho$.
\end{enumerate}

\section{Definitions}

\subsection{Notation}
\label{sec:notation} %

Throughout this paper we will denote the number of time-intervals,
transmitting antennas and receiving antennas with $T$, $M$, $N$,
respectively. $R$, $K$ and $Q$ will represent $R=\min(M,N)$,
$K=\min(T,N)$ and $Q=\max(M,N)-\min(M,N)$.

In addition, we will use bold-faced upper-case letters to
represent matrices, e.g. $\vec X$, with elements given by
$X_{ij}$, bold-faced lower-case letters for column vectors, e.g.
$\vec x$, with elements $x_i$, and non-bold lower-case letters for
scalar quantities. $\Tr{\vec X}$ will represent the trace of $\vec
X$, while the superscripts $T$ and $\dagger$ will indicate
transpose and Hermitian conjugate operations. The determinant of a
matrix will be represented by $\det(\vec X)$ or by $\det(X_{ij})$.
Also, $\vec I_n$ will denote the $n$-dimensional identity matrix,
while $\vec J_n$ will represent a $T\times T$ matrix with zeros in
all elements other than the first $n$ diagonals, which have unit
value.

The complex, circularly symmetric Gaussian distribution with
zero-mean and unit-variance will be denoted by ${\cal CN}(0,1)$.

The per-symbol normalized mutual information will be given by
$\hat{I}$, measured in bits/sec/Hz. Thus for the case of USTM,
$\hat{I}_{USTM} = I_{USTM}/T$, while for DUSTM, $\hat{I}_{DUSTM} =
I_{DUSTM}/M$.

\subsection{System Model}

We consider the case of single-user transmission from $M$ transmit
antennas to $N$ receive antennas over a narrow-band block-fading
channel. The channel coefficients are assumed to be constant over
time intervals of length $T$, after which they acquire independent
values, which in turn remain constant for the same time interval.
The received $T\times N$-dimensional complex signal $\vec X$ can
be written in terms of the $T\times M$-dimensional transmitted
complex signal $\vec \Phi$ as
\begin{equation}\label{eq:basic_channel_eq}
  \vec X = \sqrt{\frac{\rho T}{M}} \vec \Phi \vec H + \vec W
\end{equation}
where $\vec H$ is a $M\times N$ matrix with the channel
coefficients from the transmitting to the receiving arrays and
$\vec W$ is the $T\times N$ additive noise matrix. Both $\vec H$
and $\vec W$ are assumed to have elements that are independent and
${\cal CN}(0,1)$-distributed. Their instantaneous values are
assumed to be unknown to both the transmitter and the receiver.
The first term in (\ref{eq:basic_channel_eq}) is normalized, so
that $\rho$ is the total average signal-to-noise ratio (SNR)
transmitted from all antennas.

\subsection{Unitary Matrices for Isotropic and Differential USTM}

In this paper we will be dealing with unitary input distributions
$\vec \Phi$. For the case of USTM $\vec \Phi$ is a member of the
$S(M,T)$ Stiefel manifold (see \cite{Boothby_book}) i.e. the set
of all complex $T\times M$ matrices, such that
\begin{equation}\label{eq:Phiunitary}
 \vec \Phi^\dagger \vec \Phi = \vec I_M
\end{equation}
Note that it is implicitly assumed here that $T\geq M$, since only
thus can $M$ $T$-dimensional vectors be mutually orthogonal.

It is convenient here to introduce $\vec \Phi_\perp$, the
$T\times(T-M)$ orthogonal complement of $\vec \Phi$, i.e. with
\begin{equation}\label{eq:orthogonal_complement_def}%
\vec \Phi \vec \Phi^\dagger + \vec \Phi_\perp \vec
\Phi_\perp^\dagger = \vec I_T \,\,\,\, \mbox{and} \,\,\,\, \vec
\Phi_\perp^\dagger \vec \Phi_\perp = \vec I_{T-M}
\end{equation}
so that $\vec {\bar \Phi} = [\vec \Phi \, \vec \Phi_\perp ]$ is a
$T\times T$ unitary matrix with $\vec {\bar \Phi}^\dagger \vec
{\bar \Phi} = \vec {\bar \Phi} \vec {\bar \Phi}^\dagger = \vec
I_T$.

For the case of DUSTM, we restrict ourselves to the $U(M)$
subgroup of the $S(M,2M)$ Stiefel manifold, such that
\cite{Hochwald2000_DUSTM}
\begin{equation}\label{eq:Phiunitary_DUSTM}
  \vec \Phi = \frac{1}{\sqrt{2}} \left[
  \begin{array}{c} \vec I_M
  \\ \vec U
  \end{array} \right]
\end{equation}
where $\vec U$ is an $M\times M$ unitary matrix.

\subsection{Mutual Information}

For fixed $\vec \Phi$, $\vec X$ in (\ref{eq:basic_channel_eq}) is
a sum of two Gaussian matrices, therefore its probability density
conditional on $\vec \Phi$ can be written as
\begin{eqnarray}\label{eq:p(X|Phi)_def0}
  p(\vec X|\vec \Phi) &=&     \frac{\exp\left(-\Tr{ \vec X^\dagger
 \left[\vec I_{T} + \frac{\rho T}{M}
  \bf{\Phi\Phi}^\dagger \right]^{-1}
  \vec X}\right)}{\pi^{TN}\det(\vec I_T+\frac{\rho T}{M}\bf{\Phi\Phi}^\dagger)^{N}}
 \end{eqnarray}
To evaluate the inverse of the matrix in the exponent we use
(\ref{eq:Phiunitary}), \ref{eq:orthogonal_complement_def} to get
the expression $\vec \Phi\vec\Phi^\dagger = \vec {\bar \Phi} \vec
J_M \vec {\bar \Phi}^\dagger$. Applying this we get
\begin{eqnarray}\label{eq:inverse1+PhiPhi}
 \left[\vec I_{T} + \frac{\rho T}{M}
  \bf{\Phi\Phi}^\dagger \right]^{-1} &=& \left[\vec I_{T} + \frac{\rho T}{M}
  {\bar \Phi} \vec J_M \vec {\bar \Phi}^\dagger \right]^{-1} %
  \\ \nonumber %
 &=& {\bar \Phi} \left[\vec I_{T} + \frac{\rho T}{M}
   \vec J_M  \right]^{-1} \vec {\bar \Phi}^\dagger
  \\ \nonumber %
 &=& \vec \Phi_\perp \vec \Phi_\perp^\dagger + \frac{M}{M+\rho T}  \vec \Phi \vec
 \Phi^\dagger
   \\ \nonumber %
 &=& \vec I_T -\frac{\rho T}{M+\rho T}
  \bf{\Phi\Phi}^\dagger
 \end{eqnarray}
We can therefore express $p(\vec X|\vec \Phi)$ as
\begin{eqnarray}\label{eq:p(X|Phi)_def}
p(\vec X|\vec \Phi) &=&  \frac{\exp\left(-\Tr{ \vec X^\dagger
 \left[\vec I_{T} -\frac{\rho T}{M+\rho T}
  \bf{\Phi\Phi}^\dagger \right]
  \vec X}\right)}{\pi^{TN}(1+\frac{\rho T}{M})^{MN}}
\end{eqnarray}

The mutual information between $\vec X$ and $\vec \Phi$ is given
by
\begin{equation}\label{eq:I_X_Phi}
  I(\vec X;\vec \Phi) = \int d\vec \Phi \, p(\vec \Phi) \int d\vec X \, p(\vec
  X|\vec \Phi) \log_2\left(\frac{p(\vec
  X|\vec \Phi)}{p(\vec X)}\right)
\end{equation}
$p(\vec X)$ is the received signal probability density given by
\begin{equation}\label{eq:p(X)}
  p(\vec X) = \int d\vec \Phi \, p(\vec  X|\vec \Phi) \equiv \langle p(\vec  X|\vec
  \Phi)\rangle
\end{equation}
where we introduced the notation $\langle \cdot \rangle$ as the
integration over $\vec \Phi$.

The integration over $\vec \Phi$ in (\ref{eq:I_X_Phi}) can be
eliminated by noting \cite{Hassibi2002_USTM} first that
\begin{equation}\label{eq:p(X|Phi)_property}
p(\vec X|\vec \Phi) = p(\vec {\bar \Phi}^\dagger \vec X|\vec
\Phi_0)
\end{equation}
The choice of $\vec \Phi_0$ depends on the particular application.
Thus, for the case of USTM the following expression can be used
\begin{equation}\label{eq:Phi0USTM}
\vec \Phi_0 = \left[
\begin{array}{c}
\vec I_M \\ \vec 0_{T-M} \end{array} \right]
\end{equation}
while for DUSTM it is convenient to use
\begin{equation}\label{eq:Phi0DUSTM}
\vec \Phi_0 = \frac{1}{\sqrt{2}} \left[
\begin{array}{c}
\vec I_M \\ \vec I_M \end{array} \right]
\end{equation}
which is the identity matrix of matrices of the form of
(\ref{eq:Phiunitary_DUSTM}). Using (\ref{eq:p(X|Phi)_property})
and through the change of variables $\vec X \rightarrow \vec {\bar
\Phi}^\dagger \vec X $, which leaves the $\vec X$-integration
measure unaffected, we completely eliminate any non-trivial $\vec
\Phi$-dependence of the integrand of (\ref{eq:I_X_Phi}). The
remaining $\int d{\bf \Phi}p(\vec \Phi)$ can be easily integrated
to give unity and thus is disregarded. This results to
\begin{equation}\label{eq:I_X_Phi_noPhi_int}
  I(\vec X;\vec \Phi) = \int d\vec X \, p(\vec
  X|\vec \Phi_0) \log_2\left(\frac{p(\vec
  X|\vec \Phi_0)}{p(\vec X)}\right)
\end{equation}

\section{Closed-form solution of $p(\vec X)$ for DUSTM}
\label{sec:Closed-form solution pX_DUSTM} %

When dealing with DUSTM, it is convenient to express the
conditional probability in terms of $\vec U$, defined in
(\ref{eq:Phiunitary_DUSTM}). Thus, if we express $\vec X$ as
\begin{equation}\label{eq:X=X1_X2}
  \vec X = \left[ \vec X_1 \, \vec X_2 \right]^T
\end{equation}
where both $\vec X_1$ and $\vec X_2$ have dimensions $M\times N$,
then (\ref{eq:p(X|Phi)_def}) can be rewritten in terms of $\vec
X_1$, $\vec X_2$ and  $\vec U$  as
\begin{eqnarray}\label{eq:P_XU}
  p(\vec X|\vec U) &=& \frac{\exp\left(-\frac{1+\rho}{1+2\rho}\Tr{\vec X^\dagger_1 \vec
  X_1+ \vec X^\dagger_2\vec X_2}
  \right)}{\pi^{2MN}(1+2\rho)^{MN}}\\ \nonumber
 &\times& \exp\left(\frac{\rho}{1+2\rho}\Tr{ \vec
  X_2 \vec X^\dagger_1\vec U+ \vec X_1 \vec X^\dagger_2\vec U^\dagger}
  \right)
   \\ \nonumber
\end{eqnarray}
Combining this with (\ref{eq:p(X)}) we get
\begin{eqnarray}\label{eq:P_X}
  p(\vec X) &=&\frac{\exp\left(-\frac{1+\rho}{1+2\rho}\Tr{\vec X^\dagger_1 \vec
  X_1+ \vec X^\dagger_2\vec X_2}
  \right)}{\pi^{2MN}(1+2\rho)^{MN}} \\ \nonumber
  &\times &
  \left< \exp\left(\beta\Tr{\vec
  X_2 \vec X^\dagger_1 \vec U+ \vec X_1 \vec X^\dagger_2\vec U^\dagger}
  \right) \right>
\end{eqnarray}
where
\begin{equation}\label{eq:beta_def}
 \beta=\frac{\rho}{1+2\rho}
\end{equation}
We can now use the result of \cite{Brower1981_LatticeQCD} to get
\begin{eqnarray}\label{eq:ave_eqxTr}
    \left< \exp\left(\beta\Tr{ \vec
  X_2 \vec X^\dagger_1 \vec U+ \vec X_1 \vec X^\dagger_2 \vec U^\dagger}
  \right) \right> = \,\,\,\,\,\, &&  \nonumber \\
   \prod_{k=0}^{M-1} k!
  \frac{\det\left(y_j^{(i-1)/2}
  I_{i-1}(2y_j^{1/2})\right)}{\det\left(y_j^{i-1}\right)} &&
\end{eqnarray}
where $y_j$ for $j=1\ldots M$ are the eigenvalues of $\beta^2 \vec
X_1 \vec X^\dagger_2 \vec X_2 \vec X^\dagger_1$ (or the squares of
the svd's of $\beta\vec X_2 \vec X^\dagger_1$). This equation is
essentially the generating functional of $\vec U$: Any moment of
$\vec U$ can be evaluated by taking arbitrary derivatives with
respect of elements of the matrix $ \vec X_2 \vec X^\dagger_1$ on
both sides of (\ref{eq:ave_eqxTr}) and subsequently setting this
matrix to zero.

The determinant in the denominator is the Vandermonde determinant
\begin{eqnarray}
\label{eq:vandermonde_det}%
\Delta(\{y_j\}) &=&  \det (y_j^{i-1}) \nonumber \\
&=& \left|
\begin{array}{cccc}
1 & 1 & \cdots & 1 \\
y_1 & y_2 & \cdots & y_M \\
y_1^2 & y_2^2 & \cdots & y_M^2 \\
\vdots & \vdots & \cdots & \vdots \\
y_1^{M-1} & y_2^{M-1} & \cdots & y_M^{M-1}
\end{array} \right|
\end{eqnarray}
while the determinant in the numerator can be written explicitly
as
\begin{eqnarray}
\label{eq:I0_det}%
&& \det\left(y_j^{(i-1)/2}
  I_{i-1}(2y_j^{1/2})\right)  = \\ \nonumber
&& \left| \begin{array}{ccc}
I_0(2y_1^{1/2})) &  \cdot\cdot & I_0(2y_M^{1/2}) \\
y_1^{1/2}I_1(2y_1^{1/2})  &  \cdot\cdot & y_M^{1/2}I_1(2y_M^{1/2}) \\
y_1 I_2(2y_1^{1/2}) &  \cdot\cdot & y_M I_2(2y_M^{1/2}) \\
\vdots & \cdot\cdot & \vdots \\
y_1^{(M-1)/2}I_{M-1}(2y_1^{1/2}) & \cdot\cdot &
y_M^{(M-1)/2}I_{M-1}(2y_M^{1/2})
\end{array} \right|
\end{eqnarray}
where $I_n(x)$ is the modified Bessel function of order $n$.

One has to exercise caution in evaluating (\ref{eq:ave_eqxTr}) in
the case $M<N$. The reason is that only $R$ singular values of
$\vec X_1\vec X_2^\dagger$ are non-zero. Therefore, both the
determinants in the numerator and the denominator vanish. However,
the ratio remains finite. Using Lemma \ref{lemma:vandermonde} in
Appendix \ref{app:vandermonde} we can show that
\begin{eqnarray}\label{eq:ave_eqxTr_QR}
    \langle \exp\left(\beta\Tr{\vec
  X_2 \vec X^\dagger_1  \vec U+ \vec X_1 \vec X^\dagger_2 \vec U^\dagger}
  \right) \rangle = && \\ \nonumber
   \prod_{k=M-R}^{M-1} k! \,
  \frac{\det\left(y_j^{(M-R+i-1)/2}
  I_{M-R+i-1}(2y_j^{1/2})\right)}{\det\left(y_j^{M-R+i-1}\right)} &&
\end{eqnarray}
where the range of the indices in the determinants are
$i,j=1,\ldots R$.

\section{Mutual Information of DUSTM}

Using (\ref{eq:P_XU}), (\ref{eq:P_X}) and (\ref{eq:ave_eqxTr_QR})
we can now express the ratio $p(\vec X|\vec \Phi_0)/p(\vec X)$ as
\begin{eqnarray}\label{eq:LLR_DUSTM}
&&\log_2\left( \frac{p(\vec X|\vec \Phi_0)}{p(\vec X)}\right) =
\\ \nonumber %
&& \log_2\left[\frac{\prod_{k=M-R}^{M-1} \frac{1}{k!} \,\,\,
\det\left(y_j^{M-R+i-1}\right)}{\det\left(y_j^{(M-R+i-1)/2}
  {\tilde I}_{M-R+i-1}(2y_j^{1/2})\right)}\right]
 \\ \nonumber
&& + \left(\beta\Tr{\vec X^\dagger_1 \vec
  X_2+ \vec X^\dagger_2\vec X_1} - 2 \sum_{i=1}^R y_i^{1/2}
  \right) \log_2 e
\end{eqnarray}
In the above equation we have defined ${\tilde I}_n(x) =
I_n(x)e^{-x}$ and we have multiplied both numerator and
denominator of the expression inside the log with $\exp\left(-
2\sum_{i=1}^R y_i^{1/2}\right)$, so that neither will have
exponentially increasing terms for large $y_i$.

To evaluate the mutual information, (\ref{eq:LLR_DUSTM}) needs to
averaged over realizations of $\vec X_1$, $\vec X_2$, which are
generated with probability distribution $p(\vec X|\vec \Phi_0)$.
This corresponds to $\vec X_1$, $\vec X_2$ having Gaussian
correlations given by
\begin{eqnarray}
\label{eq:X1_2variances}%
E\left[X_{1i\alpha}^* X_{1j\beta}\right] &=& (1+\rho) \delta_{ij}\delta_{\alpha\beta} \\
E\left[X_{2i\alpha}^* X_{2j\beta}\right] &=& (1+\rho) \delta_{ij}\delta_{\alpha\beta} \\
E\left[X_{1i\alpha}^* X_{2j\beta}\right] &=& \rho
\delta_{ij}\delta_{\alpha\beta}
\end{eqnarray}

\section{Mutual Information of USTM}

In the next section we will compare the mutual information of
DUSTM to that of USTM. Thus, for completeness, we review here the
results obtained in \cite{Hassibi2002_USTM} regarding USTM. We
start with the conditional probability $p(\vec X|\vec \Phi)$
\begin{equation}\label{}
  p(\vec X|\vec \Phi) = \frac{\exp\left(-\Tr{ \vec X^\dagger\vec X}\right)
  \exp\left(\alpha\Tr{ \vec X^\dagger \vec \Phi \vec \Phi^\dagger \vec
  X}\right)}{\pi^{TN}(1+\rho T/M)^{MN}}
\end{equation}
where $\vec X$ is a $T\times N$ complex matrix, $\vec \Phi$ is a
$T\times N$ unitary matrix and
\begin{equation}\label{eq:alpha_def}
  \alpha = \frac{\rho T}{M+\rho T}
\end{equation}
 In \cite{Hassibi2002_USTM} the
received signal probability density was found to be
\begin{eqnarray}
\label{eq:p(X)_USTM_Marzetta}%
p(\vec X) &=& \int d\vec\Phi \, p(\vec X|\vec \Phi) \\ \nonumber
&=& \frac{\exp\left(-\Tr{\vec
X^\dagger \vec X}\right)}{\pi^{TN}\left(1+\rho T/M\right)^{MN}} \\
\nonumber &\times& %
\langle \exp\left(\alpha \Tr{{\bf X^\dagger \Phi \Phi^\dagger X}}
\right) \rangle
\end{eqnarray}
where the average over $\vec \Phi$, expressed as $\langle \cdots
\rangle$ was performed as follows:
\begin{eqnarray}
\label{eq:multi_int_iustm_def}%
&&\langle \exp\left(\alpha \Tr{{\bf X^\dagger \Phi \Phi^\dagger
X}}
\right) \rangle  \\ \nonumber %
&=&\frac{C_{TM}}{M!}\int\frac{dt_1}{2\pi}\cdots\int\frac{dt_M}{2\pi}
\\ \nonumber %
 &\times& \prod_{m=1}^M\left[\frac{e^{-it_m}}{(-\alpha
y_1-it_m)\cdots(-\alpha
y_K-it_m)(-it_m)^{T-K}}\right] \\ \nonumber %
&\times&\prod_{l<m}(-it_m-it_l)^2\\ \nonumber %
&=& C_{TM}\left|\det \vec F\right|
\end{eqnarray}
where the constant $C_{TM}$ is equal to
\begin{equation}\label{eq:C_TM_def}
  C_{TM} = \frac{(T-1)!\cdots(T-M)!}{(M-1)!\cdots(0)!}
\end{equation}
and $\vec F$ is a $M\times M$ Hankel matrix with entries given by
\begin{eqnarray}
\label{eq:Fmn_USTM_Marzetta}%
F_{mn} &=& \sum_{k=1}^K \frac{e^{\alpha y_k}}{(\alpha
y_k)^q\prod_{l\neq k} (\alpha y_k-\alpha y_l)} \\ \nonumber
&\times& \left\{ \begin{array}{cc} %
 \frac{\gamma(q,\alpha y_k)}{\Gamma(q)}, & q\geq 1 \\ 1, &
q\leq 0
\end{array} \right.
\end{eqnarray}
In the above expression, $q = T-K-m-n+2$, $\gamma(n,x)$ is the
incomplete $\Gamma$ function and $y_n$, for $n=1,\ldots,K$ are the
non-zero eigenvalues of the $N\times N$ matrix ${\bf X^\dagger
X}$. As in the case of DUSTM, to numerically calculate the mutual
information one needs to average the log-ratio $\log_2(p(\vec
X|\vec \Phi_0)/p(\vec X))$, where $\vec \Phi_0$ is given by
(\ref{eq:Phi0USTM}), with respect to $\vec X$, which has
probability density $p(\vec X|\vec \Phi_0)$. It is convenient to
write ${\bf X^\dagger X}$ as
\begin{equation}\label{eq:iustm_XX_stat}
  {\bf X^\dagger X} = \left(1+\frac{\rho T}{M}\right) {\bf
  X_1^\dagger X_1} + {\bf X_2^\dagger X_2}
\end{equation}
where $\vec X_1$, $\vec X_2$ are $M\times N$, $(T-M)\times N$
complex Gaussian, unit-variance matrices.

\section{Analysis and Comparison to Isotropic USTM}

In section \ref{sec:Numerical Simulations} below, we present
numerical results on the mutual information of DUSTM and compare
them to corresponding USTM results. However, before that, it is
instructive to analyze the asymptotic behavior of the mutual
information in both small and large SNR regimes. As we shall see,
this exact asymptotic analysis of both USTM and DUSTM will provide
insight and quantitative agreement with numerical simulations.

\subsection{Low $\rho$ region}

To obtain the small $\rho$ behavior we expand the exponent in the
log-ratio $\log_2\left(p(\vec X|\vec \Phi_0)/p(\vec X)\right)$ and
integrate over the fields. For the DUSTM case in Appendix
\ref{app:small_rho_forDUSTM} we obtain
\begin{equation}%
\label{eq:DUSTM_small_rho}
\hat{I}_{DUSTM} \approx \rho^2
N\left[1-2\rho+\frac{\rho^2}{2}\left(5-\frac{N}{M}\right)\right]\log_2e
\end{equation}
For small small $\rho$, we see that $\hat{I}_{DUSTM}$ is an
increasing function of $M$. As a result, under the constraint of
the channel being constant over $T$ time-intervals, the optimal
number of transmitting antennas is $M_{opt} = T/2$.

For comparison, in Appendix \ref{app:small_rho_forUSTM} we
calculate the mutual information for USTM for small $\rho$. The
final result up  to ${\cal O}(\rho^3)$ is
\begin{eqnarray}\label{eq:USTM_small_rho}
  \hat{I}_{USTM} &\approx& \frac{N
\rho^2}{2M}(T-M)\left[1-\frac{2\rho
T}{3M}\left(1+\frac{M}{T}\right)\right]\log_2e  \nonumber \\
 &=& \frac{\rho^2 N}{2}(1-2\rho)\log_2e
\end{eqnarray}
where the last equality holds for $T=2M$. We see that for $T=2M$,
$\hat{I}_{DUSTM}\approx 2\hat{I}_{USTM}$ up to order ${\cal
O}(\rho^3)$! Also, for fixed $T$ and $N$, $\hat{I}_{USTM}$ is
actually a decreasing function of the number of transmitting
antennas $M$, with optimal $M=1$. This can be seen in Fig.
\ref{fig:D_IUST_capacity}, where the optimal $M$ at low $\rho$ is
1.

It is important to note that for $\rho\ll 1$, the mutual
information for both schemes scales as $\rho^2$, rather than
$\rho$ as in the coherent case. This behavior has been pointed out
by \cite{Rao2004_LowSNRAnalysisMIMOlinks,
Prelov2004_2ndorderAsymptotics}. Thus, at small SNR, the lack of
knowledge of the channel becomes increasingly problematic. This is
generally the case for unitary space-time modulated schemes.

\subsection{High $\rho$ region}

In Appendix \ref{app:large_rho_forDUSTM} we obtain the large
$\rho$ behavior of the mutual information of DUSTM, which to
${\cal O}(\log_2\rho/\rho)$ is
\begin{eqnarray}\label{eq:DUSTM_large_rho}
  \hat{I}_{DUSTM} &=& \frac{1}{M} \left[
  R\left(M-\frac{R}{2}\right)\log_2\rho  +  A_{MN}\right]\\ \nonumber
  &+& {\cal
  O}\left(\frac{\log_2\rho}{\rho}\right)
\end{eqnarray}
where
\begin{eqnarray}\label{eq:AMN_DUSTM}
  A_{MN} &=& \frac{R}{2}\log_2 (4\pi)
  -R\left(M-\frac{R}{2}\right)\log_2(2e)
  \nonumber \\
  &-& \sum_{k=M-R}^{M-1} \log_2 k! + R\left(M-R+\frac{1}{2}\right) {\cal L}_1(M,N)
  \nonumber \\
   &+& \frac{1}{2}R(R-1) {\cal L}_2(M,N)
\end{eqnarray}
is a constant, independent of $\rho$. In (\ref{eq:AMN_DUSTM}) we
have defined the quantities ${\cal
L}_1(M,N)=E\left[\log_2\lambda_1\right]$ and ${\cal
L}_2(M,N)=E\left[\log_2(\lambda_1+\lambda_2)\right]$, where
$\lambda_{1,2}$ are distinct non-zero singular values of an
$M\times N$ matrix with independent ${\cal CN}(0,1)$ entries.
 Their explicit expressions are given in (\ref{eq:ave_log_lambda}), (87). %

Similarly, in Appendix \ref{app:large_rho_forUSTM} we derive the
asymptotic large-$\rho$ form of the mutual information for USTM
(for $T\geq M$)
\begin{eqnarray}\label{eq:USTM_large_rho}
  \hat{I}_{USTM} &=& \frac{1}{T}\left[R(T-M)\log_2\rho +
  B_{TMN}\right] \\ \nonumber
  &+& {\cal O}\left(\frac{\log_2\rho}{\rho}\right)
\end{eqnarray}
with
\begin{eqnarray}\label{eq:BTMN_USTM_MleqN}%
  B_{TMN} &=& R(T-M)\left(\log_2\frac{T}{Me} + {\cal L}_1(M,N)\right) \\ \nonumber
   &-& \log C_{TM}- \log_2 \left|\det\vec G\right|
\end{eqnarray}
with ${\cal L}_1(M,N)$ given in (\ref{eq:ave_log_lambda}). The
last term appears only for $M<N$ and the elements of $\vec G$ are
given in (\ref{eq:G_matrix_large_rho_iusmt}). It is important to
note that for $T=M$ the mutual information vanishes to the order
calculated above, since in that case the mutual information is
identically zero.

The leading terms, proportional to $\log_2\rho$ in
(\ref{eq:DUSTM_large_rho}) and (\ref{eq:USTM_large_rho}) provide
insight on the large $\rho$ behavior of DUSTM and USTM. Starting
with (\ref{eq:DUSTM_large_rho}), we find that for fixed $N$, the
mutual information $\hat{I}_{DUSTM}$ is an increasing function of
$M$. Thus, as we found in the small $\rho$ case in the previous
section, to maximize the mutual information, one should use the
maximum number of transmitting antennas consistent with the
constraint that the channel is constant over $2M$ time-intervals.

In the case of USTM we find that, for $T>2N$ the optimal
transmitting antenna number is $M_{opt}=N$, while in the opposite
case $T\leq 2N$, the leading term is optimized for $M_{opt}=T/2$.

Once optimized over $M$, the leading terms of both
(\ref{eq:DUSTM_large_rho}) and (\ref{eq:USTM_large_rho}) are
identical to \ref{eq:cap_large_rho_incoherent}). Thus, to leading
order in $\rho$, both DUSTM and USTM are capacity achieving
schemes. Comparing the next-to-leading $\rho$-independent terms in
(\ref{eq:DUSTM_large_rho}), (\ref{eq:USTM_large_rho}) we find
that, after optimizing over $M$, the mutual information of DUSTM
is larger than that of USTM. This can be seen in Fig.
\ref{fig:D_IUST_capacity}, where the optimized-over-$M$
$\hat{I}_{DUSTM}$ and $\hat{I}_{USTM}$ of
(\ref{eq:DUSTM_large_rho}) and (\ref{eq:USTM_large_rho}) are
plotted (dashed lines). This may come as a surprise if one takes
into account that for $T=2M$, the manifold of constellations used
for DUSTM (\ref{eq:Phiunitary_DUSTM}) is a subgroup of those used
in USTM. However, one should take into account that in DUSTM,
although information is sent over $M$ time-intervals, the receiver
exploits the side information that the channel has not changed
over the previous $M$ time-intervals.

\subsection{Numerical Simulations}
\label{sec:Numerical Simulations}

We now discuss the numerical simulations performed to evaluate the
mutual information for USTM and DUSTM. The simulation procedure
consists of the following steps: First we generate $L$ instances
of Gaussian complex random matrices with covariance given by
(\ref{eq:X1_2variances}) and (\ref{eq:iustm_XX_stat}) for the
DUSTM and USTM cases. For each matrix instantiation we calculate
the singular values and then we apply them to evaluate the
log-ratio $\log_2(p(\vec X|\vec \Phi)/p(\vec X))$, which we then
average over its $L$ values. For intermediate and large $\rho$ we
have found that $L\approx 4-5\cdot 10^4$ are sufficient. However,
for smaller $\rho$, at least $L=5\cdot 10^5$ are required. The
reason is that the mutual information, being ${\cal O}(\rho^2)$,
is quite small and therefore fluctuations have a more pronounced
effect.

\begin{figure*}
\centerline{%
\subfigure[$\rho = -6.0dB$]
{\epsfxsize=.50\textwidth
 \epsffile{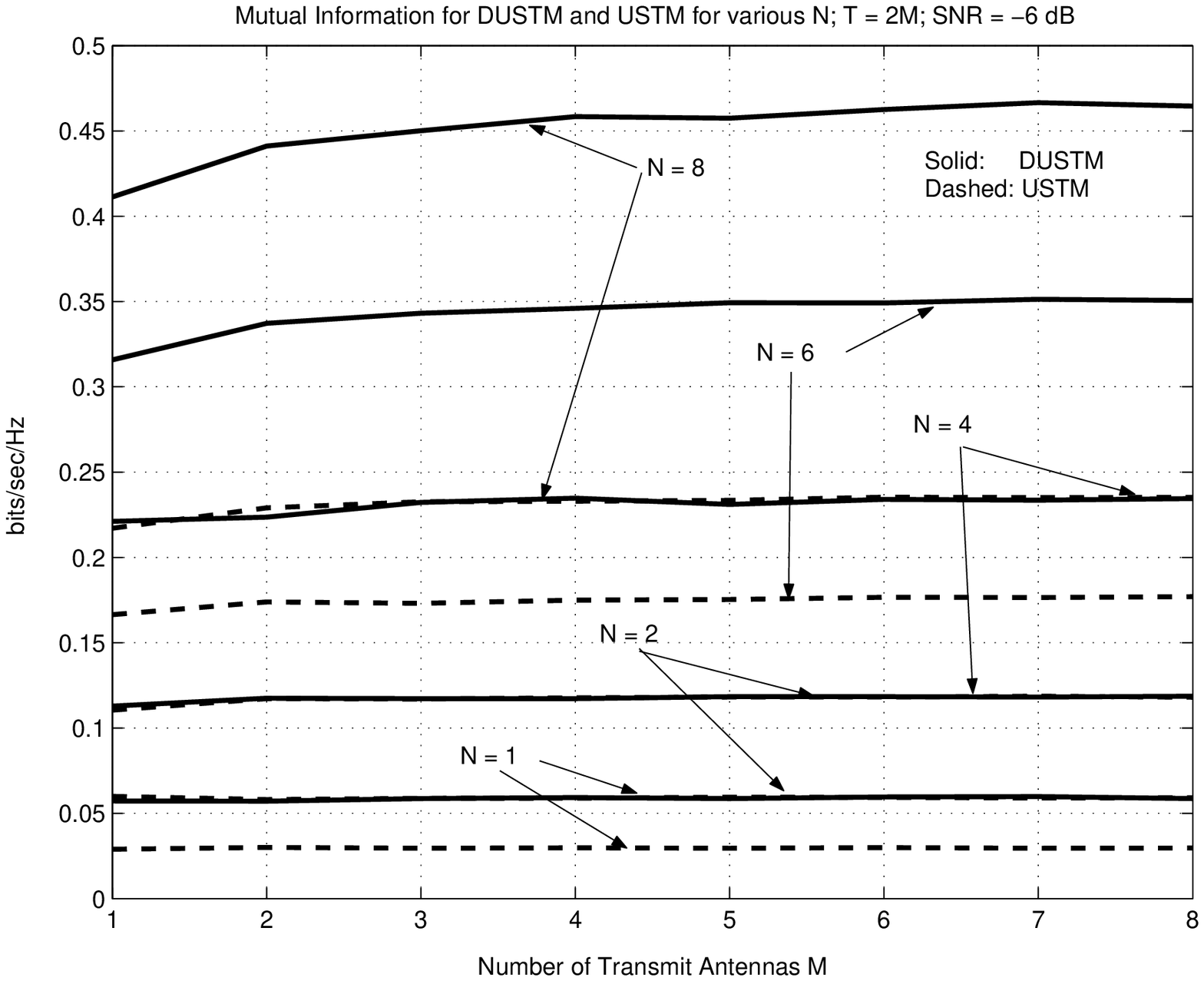} }
\hfil %
\subfigure[$\rho = 6.0dB$]
{\epsfxsize=.50\textwidth
\epsffile{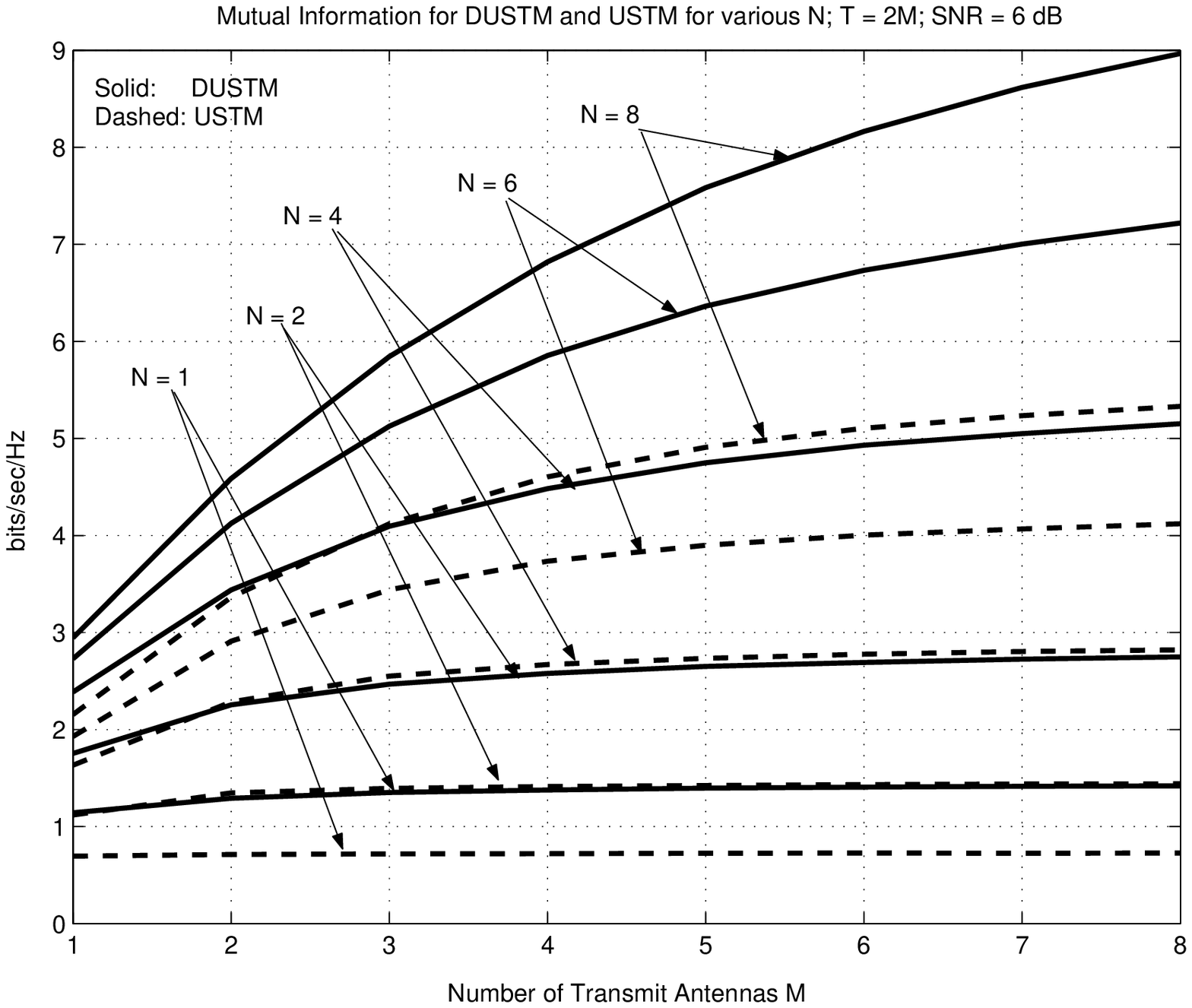} }}%
\vfil %
\centerline{%
\subfigure[$\rho = 18.0dB$]
{\epsfxsize=.50\textwidth
 \epsffile{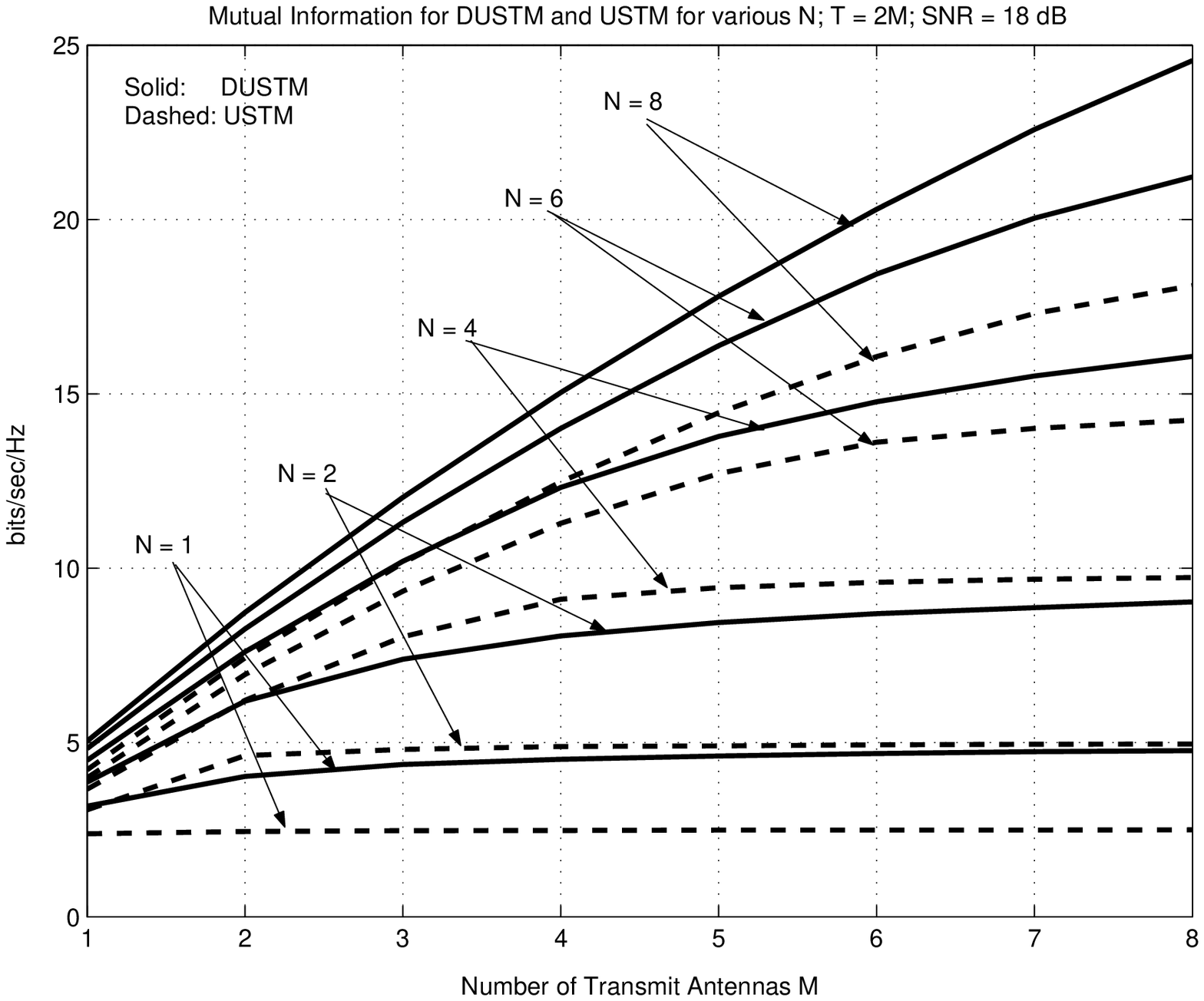} } }
\caption{Plot of normalized mutual information $\hat{I}_{DUSTM}$
(solid lines) and $\hat{I}_{USTM}$ (dashed lines) as a function of
the number of transmit antennas $M$ for different receive antenna
numbers $N$ and three SNR levels $\rho$. The coherence interval
$T$ is chosen to be $T=2M$ for proper comparison.}
\label{fig:D_IUST_mut_info}
\end{figure*}

In Fig. \ref{fig:D_IUST_mut_info} we compare the numerically
evaluated mutual information of USTM and DUSTM for low,
intermediate and relatively large SNR values. We find that for
small $\rho=-6dB$ the normalized mutual information
$\hat{I}_{DUSTM}$ is nearly exactly twice $\hat{I}_{USTM}$. This
is in agreement with (\ref{eq:DUSTM_small_rho}) and
(\ref{eq:USTM_small_rho}). Even for intermediate SNR, $\rho=6dB$
we find the approximate relation $\hat{I}_{USTM}(T=2M,M,2N)\approx
\hat{I}_{DUSTM}(T=2M,M,N)$. This approximation breaks down for
larger $\rho$.

\begin{figure*}
\centerline{%
\subfigure[$\hat{I}_{DUST}(T=2M,M,N)/\hat{I}_{IUST}(T=2M,M,N)$]
{\epsfxsize=.49\textwidth
 \epsffile{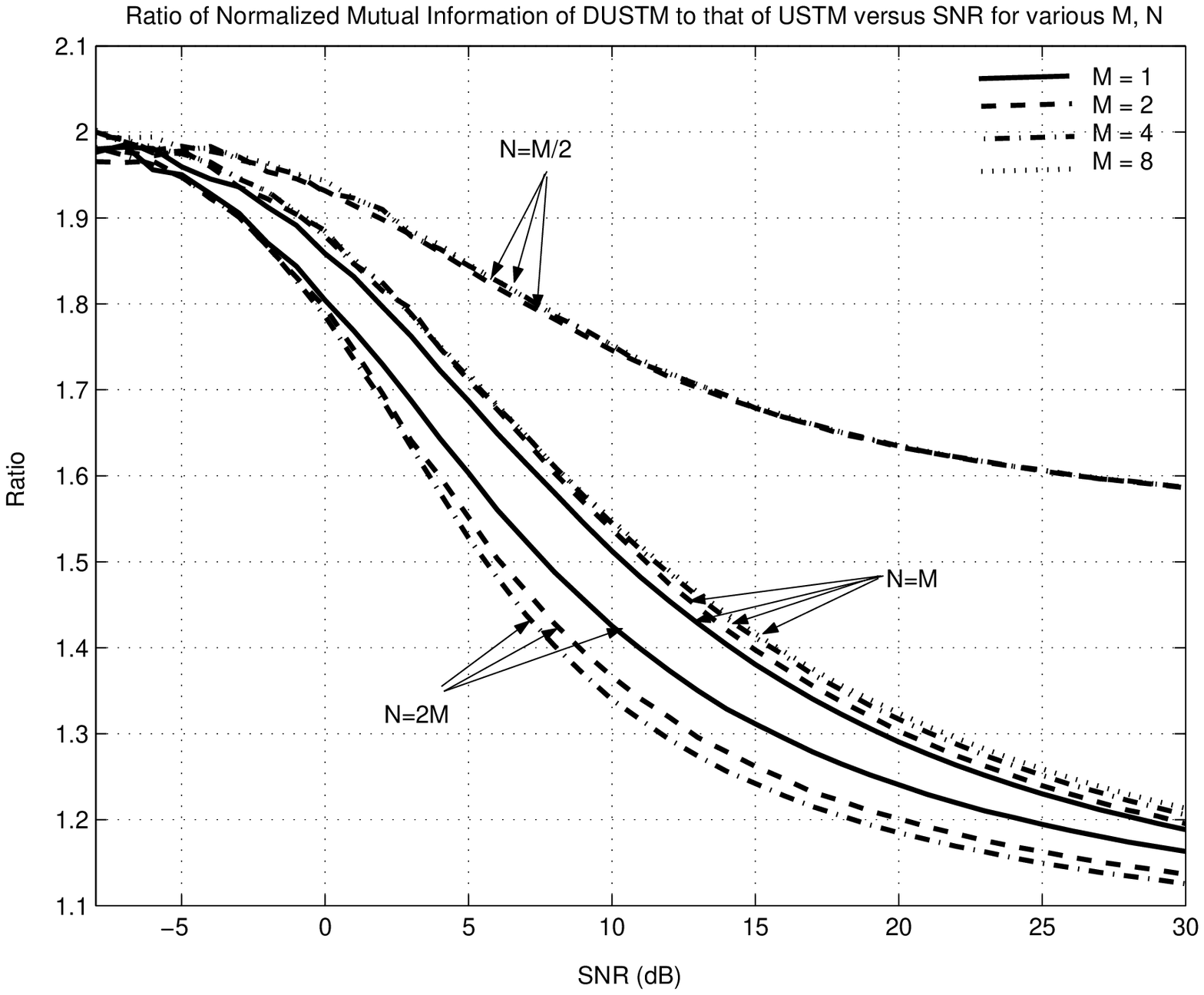} }
\hfil %
\subfigure[$\hat{I}_{DUST}(T=2M,M,N)/(M\hat{I}_{DUST}(T=2,1,1))$]
{\epsfxsize=.49\textwidth
 \epsffile{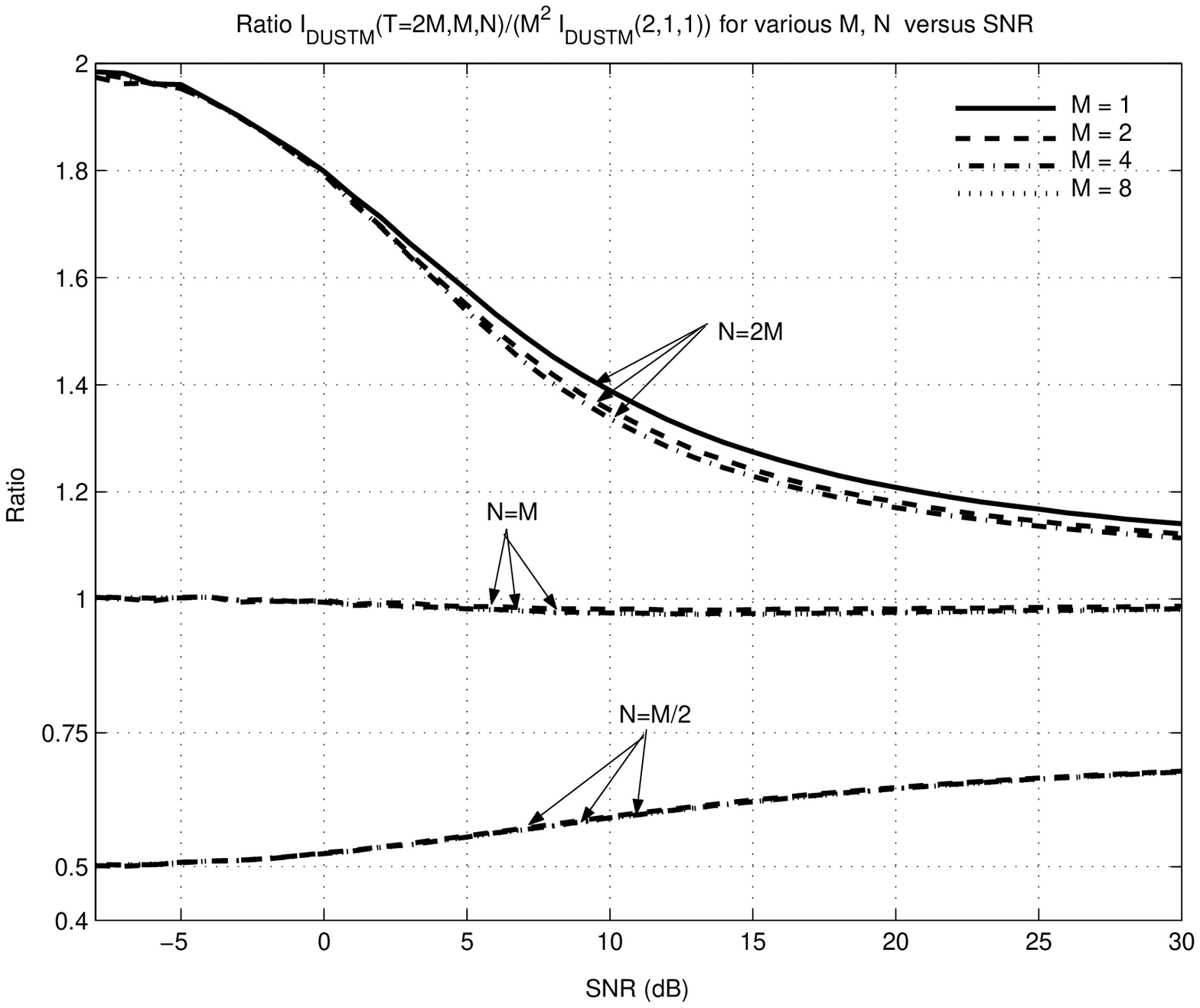}}} 
\caption{(a) Plot of the ratio between the mutual information per
symbol for the Differential USTM and the Isotropic USTM (with
$T=2M$) as a function of SNR. For low SNR, the ratio approaches 2,
as seen in the previous figure and in agreement with
(\ref{eq:DUSTM_small_rho}), (\ref{eq:USTM_small_rho}). For large
SNR the ratio appears to approach 1 (for $M=N$ and $M=N/2$) and
1.5 (for $M=2N$), as predicted from (\ref{eq:DUSTM_large_rho}),
(\ref{eq:USTM_large_rho}). For $M=N$ and $M=2N$, the ratio appears
not to depend on the number of antennas for any intermediate
$\rho$. (b) Plot of ratio between the mutual information per
symbol for the Differential USTM for various antenna numbers to
that for $M=N=1$.} \label{fig:I_DUST_ratio_vs_SNR}
\end{figure*}

Motivated by these ratio dependencies and scaling relations, in
Fig. \ref{fig:I_DUST_ratio_vs_SNR} we analyze the dependence of
ratios of $\hat{I}_{DUSTM}$ and $\hat{I}_{USTM}$ on SNR. In Fig.
\ref{fig:I_DUST_ratio_vs_SNR}(a) we plot the ratio
$\hat{I}_{DUSTM}(T=2M,M,N=rM)/\hat{I}_{USTM}(T=2M,M,N=rM)$ as a
function of $\rho$ for various values of $M$ and for $r=1/2$,
$r=1$ and $r=2$. We find that for fixed $r$, the ratios fall close
(but not on top) to each other. Their value starts from very close
to 2, for small $\rho$ and in accordance with
(\ref{eq:DUSTM_small_rho}), (\ref{eq:USTM_small_rho}), and
approaches $2(1-0.5\min(1,r))$, in agreement with
(\ref{eq:DUSTM_large_rho}), (\ref{eq:USTM_large_rho}). We note
however the slow convergence to their asymptotic values for large
$\rho$, which can be explained by the fact that both mutual
informations increase only logarithmically with $\rho$. The
closeness of the curves for fixed $r$ indicates that the ratio has
weak dependence on the actual values of $T,\,M,\,N$. Thus a
large-$T,\,M,\,N$ analysis is expected to give good results even
for small antenna numbers.

In Fig. \ref{fig:I_DUST_ratio_vs_SNR}(b) we plot the ratios
$\hat{I}_{DUSTM}(T=2M,M,N=rM)/(M\hat{I}_{DUSTM}(2,1,N=1)$ as a
function of $\rho$ for various values of $M$ and $r$.

\begin{figure*}
\centerline{%
\subfigure[Capacity--small $\rho$ region]
{\epsfxsize=.49\textwidth
 \epsffile{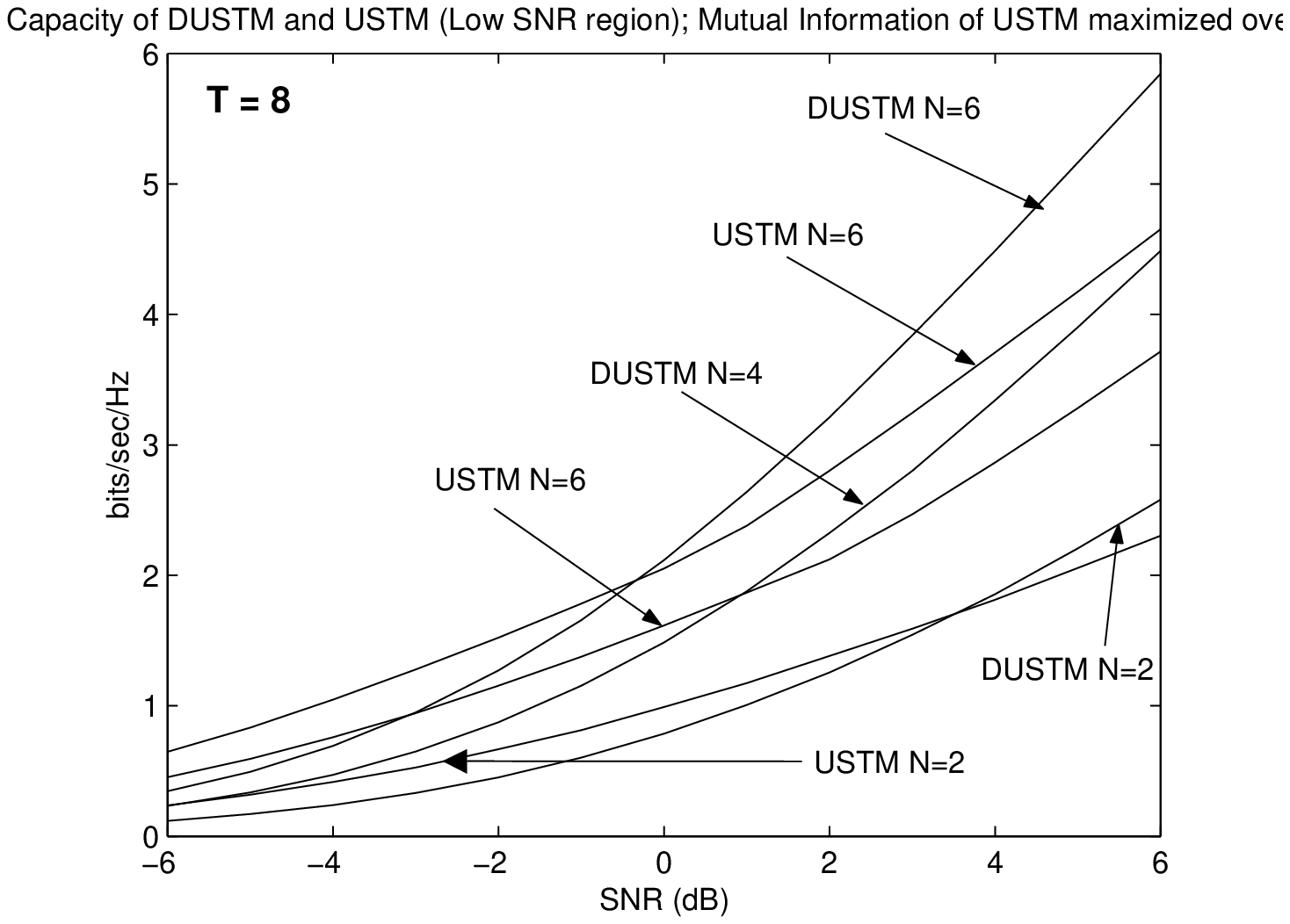} }
\subfigure[Capacity--full region]
{\epsfxsize=.49\textwidth
 \epsffile{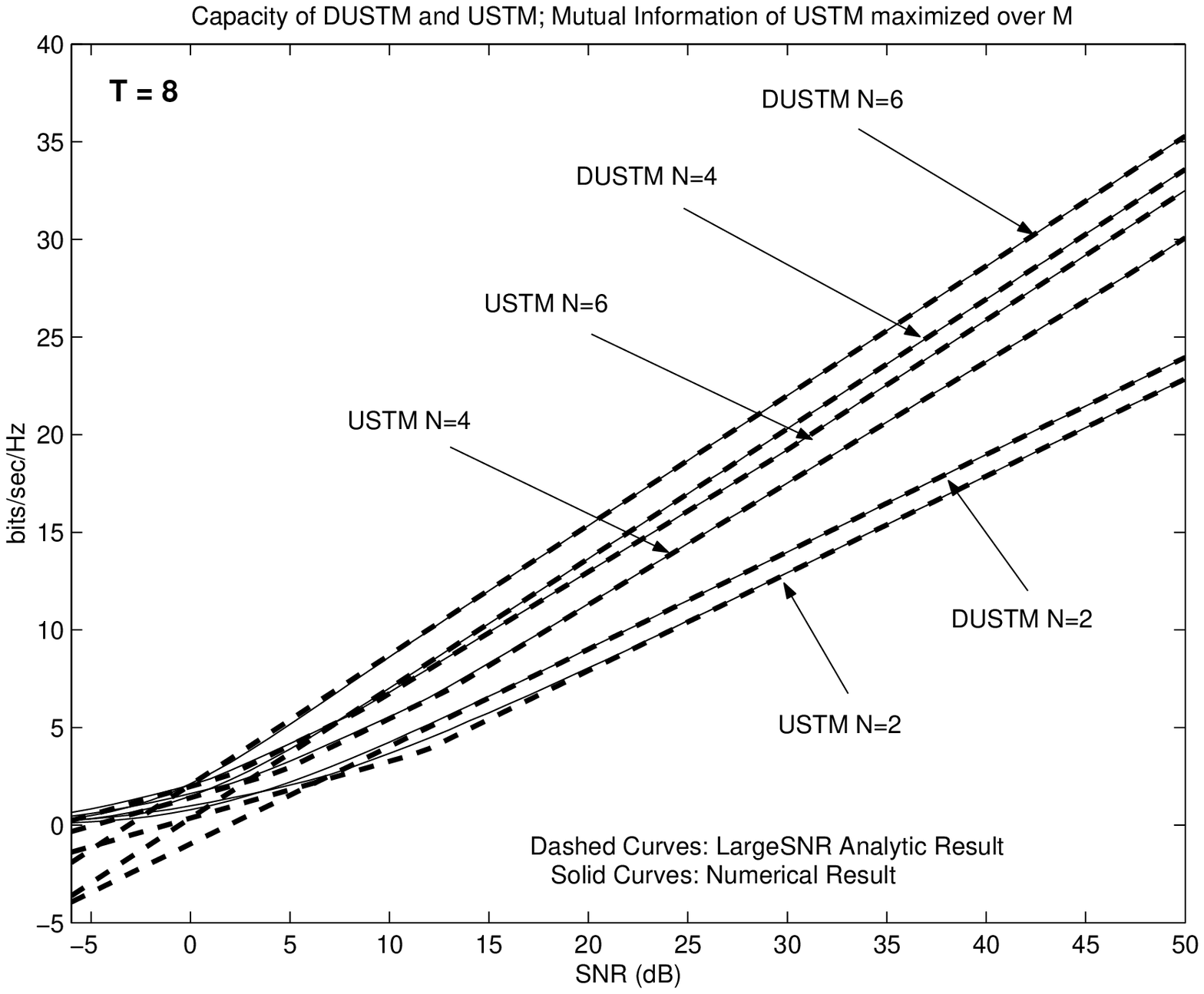} } }%
\vfil %
\centerline{%
\subfigure[Optimal number of M for USTM]
{\epsfxsize=.49\textwidth
\epsffile{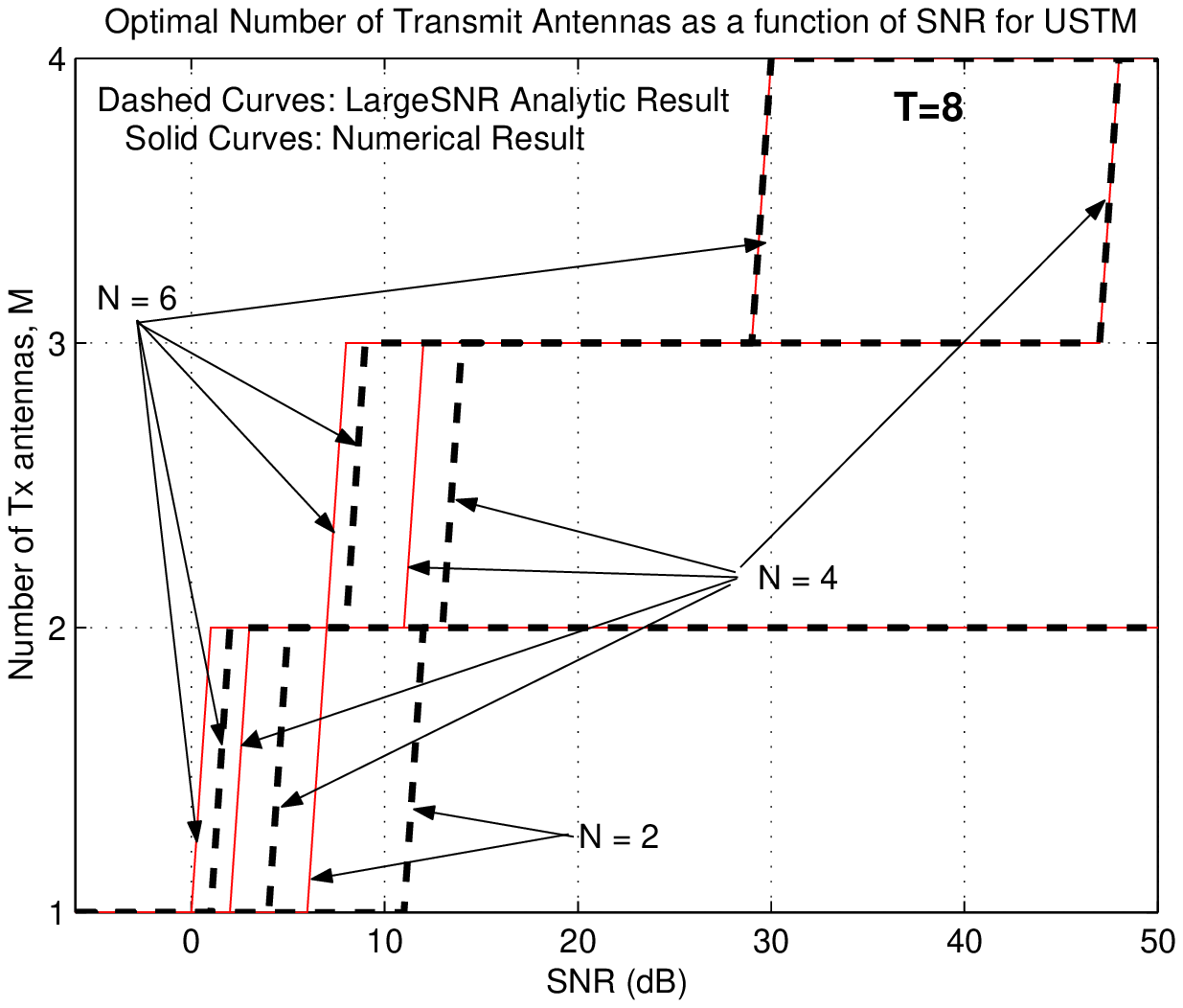} }} \caption{Plot of capacity of
DUSTM and USTM as a function of $\rho$. Here $T$ is fixed to
$T=8$. Fig. \ref{fig:D_IUST_capacity}(a) depicts the small-$\rho$
region, while Fig. \ref{fig:D_IUST_capacity}(b) shows the full
region of $\rho$-values simulated. Fig.
\ref{fig:D_IUST_capacity}(c) represents the optimal number of
transmitting antennas for USTM as a function of SNR and for
various values of $N$. The dashed curves in Fig.
\ref{fig:D_IUST_capacity}(b),(c) correspond to the analytic
expressions valid for large $\rho$ (see
(\ref{eq:DUSTM_large_rho}), (\ref{eq:USTM_large_rho})).}
\label{fig:D_IUST_capacity}
\end{figure*}

In Fig. \ref{fig:D_IUST_capacity} we analyze the mutual
information of DUSTM and USTM optimized over the number of
transmitting antennas $M$ with $T$ fixed to $T=8$ and for various
values of $N$. In Figs \ref{fig:D_IUST_capacity}(a),(b) we plot
the capacity of DUSTM and USTM defined as
\begin{eqnarray}
\label{eq:maxMIdustm} %
C_{DUSTM} &=& \max_{M^* \leq T/2} \hat{I}_{DUSTM}(T^*=2M^*,M^*,N)
\\ \nonumber
&=& \hat{I}_{DUSTM}(T,T/2,N) \\
\label{eq:maxMIustm} %
C_{USTM} &=& \max_{M^*} \hat{I}_{USTM}(T,M^*,N)
\end{eqnarray}
as a function of $\rho$ (solid curves). In Fig.
\ref{fig:D_IUST_capacity}(c) the solid curves depict the optimal
number of $M$ that maximizes $\hat{I}_{USTM}(T,M,N)$
\begin{equation}
\label{eq:argmaxMIustm} %
M_{opt} =  \arg\max_{M^*} \hat{I}_{USTM}(T,M^*,N)
\end{equation}
as a function of $\rho$. As seen in (\ref{eq:maxMIdustm}), the
optimal $M$ for DUSTM is always equal to $M=T/2$, consistent with
both low and large $\rho$ analysis. In Figs.
\ref{fig:D_IUST_capacity}(b),(c) the dashed curves represent the
capacity and optimal $M$ values as evaluated using the
large-$\rho$ asymptotic expressions of (\ref{eq:DUSTM_large_rho}),
(\ref{eq:USTM_large_rho}). Very good agreement with the exact
values (solid curves) can be seen down to moderate SNR. However,
one should note, that even though (\ref{eq:USTM_large_rho})
describes the capacity accurately down to moderate SNR, the
large-$\rho$ optimal value of $M$ as predicted by simply
maximizing the $\log\rho$ term in
(\ref{eq:cap_large_rho_incoherent})
\cite{Zheng2002_CommunicationOnGrassmannManifold} and in
(\ref{eq:USTM_large_rho}), actually becomes optimal at very large
$\rho\sim 50dB$.

Turning now to  Fig. \ref{fig:D_IUST_capacity}(a) we see that at
relatively small SNR,  $C_{USTM}$ and $C_{DUSTM}$ actually cross
each other. At high SNR, DUSTM consistently performs better than
USTM. At low SNR, USTM, when optimized over $M$ performs better
than DUSTM. This can be explained by looking at the leading term
of (\ref{eq:USTM_small_rho}): the optimal $M$ is $M_{opt} = 1$ and
$\hat{I}_{USTM}(T,1,N)$ can be higher than
$\hat{I}_{DUSTM}(T,T/2,N)$. Interestingly, the analytic estimates
at low SNR do not match very accurately to the behavior at  $\rho
\approx -6dB$.

\section{Conclusions}
\label{sec:conclusions}

In conclusion, we have found a closed-form expression for the
probability density of the received signal for differential
unitary space-time modulated (DUSTM) signals. This allowed us to
evaluate numerically the corresponding mutual information. In
addition, we calculated analytically the asymptotic form of the
mutual information for DUSTM and  USTM  for small and large SNR's.
At low SNR's the nondifferential form of USTM can outperform the
differential form if the number of transmit antennas is optimized.
However, at high enough SNR's the differential USTM outperforms
its nondifferential counterpart with respect to mutual
information. An additional advantage of DUSTM over USTM is its
simplicity of decoding, though recent progress has been reported
for decoding of nondifferential USTM
\cite{Hassibi2002_USTM_Cayley}. This suggests that DUSTM is a
promising type of transmission for non-coherent MIMO channels. It
would be interesting to test the competitive advantage of
differential USTM in cases when $T>2M$, for example when $T$ is a
higher multiple of $2M$. In that case the successive use of
differential USTM could be assessed.

\section*{Acknowledgments} We thank the anonymous reviewers and
C. Peel for useful comments.

\appendices
\section{}\label{app:vandermonde}%
\begin{lemma}
\label{lemma:vandermonde}
Let $f_j(x_i)$ represent the $i,j$-th
element of a $T\times T$ dimensional matrix. Here $f_j(x)$ for
$j=1,\ldots,T$ is a family of analytic functions and $\{ x_i\} $
is a $T$-plet of real numbers. For simplicity we represent this
matrix in terms of its columns denoted by $\vec f(x_i) =
\left[f_1(x_i)\, f_2(x_i)\, \ldots \, f_T(x_i)\right]^T$. Also we
denote by $\Delta(\{x_j\})$ the Vandermonde determinant of the
$x_j$'s
\begin{equation}\label{eq:Lemma_vander_det}
  \Delta(\{x_i\}) = \det(x_i^{(j-1)}) = \prod_{j>i} (x_j - x_i)
\end{equation}
Thus, in the limit that a  subset of $k$ members of the $T$-plet
are equal with each other (i.e. $x_1=\ldots = x_k$, for $k\leq
T$), then the ratio of $\det f_i(x_j)/\Delta(\{x_i\})$ exists and
is equal to
\begin{eqnarray}\label{eq:Lemma_vander_statement}
&&  \lim_{x_i\rightarrow x_1\, i=2,\ldots, k}
  \frac{\det\left[\vec f(x_1) \, \vec f(x_2) \, \ldots \, \vec f(x_T)\right]}{\prod_{j=1}^T \prod_{i>j}(x_i-x_j)}
  =
  \\ \nonumber
&& \frac{\det\left[\vec f(x_1) \, \vec f^{(1)}(x_1) \, \ldots \,
\vec f^{(k-1)}(x_1) \,
                  \vec f(x_{k+1}) \, \ldots \, \vec f(x_T)\right]}{\prod_{p=0}^{k-1} p!
                  \left[\prod_{i>j,\,j=k+1}^T
                  (x_i-x_j)\right] \prod_{m=k+1}^T
                  (x_m-x_1)^k} 
\end{eqnarray}
where $\vec f^{(n)}(x) $ denotes the $n$-th derivative of each of
the elements of the vector $\vec f(x)$ evaluated at $x$.
\end{lemma}
\begin{proof}
This can be proved by successively applying the l'Hospital limit
$k-1$ times on the numerator and denominator of
(\ref{eq:Lemma_vander_statement}). For the $p$th application of
this rule ($p< k$) we calculate the limit of $x_{p+1}\rightarrow
x_1$. For example, if $k=2$, both numerator and denominator in
(\ref{eq:Lemma_vander_statement}) have a simple zero in the limit
$x_2\rightarrow x_1$. Therefore, by taking a single derivative of
both and setting $x_2=x_1$ in the result gives the correct answer.
For $k=3$, we first take the $x_2\rightarrow x_1$ limit as above
and then we take the limit $x_3\rightarrow x_1$. Now both top and
bottom expressions of the ratio in
(\ref{eq:Lemma_vander_statement}) go to zero quadratically in
$(x_3-x_1)$. Hence one has to take the second derivative with
respect to $x_3$ on both top and bottom expressions. For a full
proof see \cite{Simon2004_CapacityOfCorrRandomWishartMatrices}
\end{proof}

\section{Small $\rho$ Analysis}
\label{app:small_rho}

In this section we will calculate the first four terms in the
Taylor expansion in $\rho$ of the mutual information for both the
differential and the isotropic USTM cases. The mutual information
$I(\vec X;\vec \Phi)$, (\ref{eq:I_X_Phi_noPhi_int}) can be
rewritten as
\begin{equation}
\label{eq:IXPhi_app_smallrho}%
I(\vec X;\vec \Phi) = \int d\vec X \, p(\vec
  X|\vec \Phi_0) \log_2 \frac{p(\vec  X|\vec \Phi_0)}{<p(\vec  X|\vec \Phi)>}
\end{equation}
where $\langle \cdot \rangle$ denotes average over $\vec \Phi$.
The simplest way to proceed is to expand both logarithms in powers
of $\rho$ and, where convenient, interchange the integration over
$\vec X$ and $\vec \Phi$. The expectation over $\vec X$ will be
denoted by $E[\cdot]$.

\subsection{Differential USTM} \label{app:small_rho_forDUSTM}

In the case of DUSTM, we see that by taking the ratio of $p(\vec
X|\vec U=\vec I_M)$ in (\ref{eq:P_XU}) and $p(\vec X)$ in
(\ref{eq:P_X}) the mutual information can be written as
\begin{eqnarray}
\label{eq:I_small_rho_dustm_1}%
I &=& \beta E\left[ \Tr{\vec X_2\vec X_1^\dagger  + \vec X_1 \vec
X_2^\dagger}\right]\log_2e \\ \nonumber %
& - & E\left[\log_2 \langle\exp\left(\beta \Tr{ \vec X_2 \vec
X_1^\dagger\vec U +  \vec X_1 \vec X_2^\dagger \vec U^\dagger}
\right) \rangle \right]
\end{eqnarray}
Since $\vec X_1$, $\vec X_2$ are zero-mean Gaussian quantities, we
only need to specify their variances given by
(\ref{eq:X1_2variances}). As a result, the first term in
(\ref{eq:I_small_rho_dustm_1}) can be easily evaluated to give
\begin{equation}\label{eq:aveX1X2}
\beta E\left[ \Tr{ \vec X_2 \vec X_1^\dagger+ \vec X_1\vec
X_2^\dagger }\right]= 2\beta\rho MN
\end{equation}

To deal with the second term in (\ref{eq:I_small_rho_dustm_1}), we
also need the following identities for the averages over $\vec U$.
\begin{eqnarray}
\label{eq:U_2moment}%
\langle U_{ij} U_{lk}^* \rangle &=& \frac{1}{M} \delta_{il}\delta_{jk} \\
\label{eq:U_odd_moments}%
\langle \prod_{q=1}^{2d+1}  U_{i_q j_q} \rangle &=& 0
\end{eqnarray}
where $r=0,1,\cdots$. Note that, since $\vec U$ is an element of
$U(M)$, all odd moments vanish. However, even moments other than
the second one are not easy to evaluate. In fact, even using the
simple-looking form of (\ref{eq:ave_eqxTr}) does not simplify
matters too much.

To expand the exponent of the second term in
(\ref{eq:I_small_rho_dustm_1}) in powers of $\beta$ we use the
notation
\begin{equation}\label{eq:An_notation}
  A_n =  \left(\Tr{\vec X_2 \vec X_1^\dagger \vec U + \vec X_1 \vec X_2^\dagger \vec
  U^\dagger}\right)^n
\end{equation}
We see that due to (\ref{eq:U_odd_moments}), all odd terms vanish,
$\langle A_{2r+1}\rangle = 0$. Thus, to $4^{th}$ order in $\beta$,
(\ref{eq:I_small_rho_dustm_1}) can be written as
\begin{eqnarray}
\label{eq:I_small_rho_dustm_2}%
I &=& \left( 2\beta \rho MN- %
\frac{\beta^2}{2} E\left[\langle A_2\rangle\right] \right.
\\ \nonumber &-& \left. %
\frac{\beta^4}{24} \left(E\left[\langle
A_4\rangle\right] -
3 E\left[\langle A_2\rangle^2\right] \right) %
\right)\log_2e \nonumber %
\end{eqnarray}
From (\ref{eq:U_2moment}) we get
\begin{equation}\label{eq:A2_Uave}
  <A_2> = \frac{2}{M} \Tr{ \vec X_2 \vec X_1^\dagger \vec X_1\vec X_2^\dagger}
\end{equation}
which results to $E\left[\langle A_2 \rangle \right] = \rho^2 M
N^2 + (1+\rho)^2MN^2$. Since $\beta = \rho/(1+2\rho)$ is ${\cal
O}(\rho)$ for small $\rho$, we only need to evaluate the averages
involving $A_4$ and $A_2^2$ to leading order in $\rho$, i.e. to
${\cal O}(1)$. Thus, we may neglect the ${\cal O}(\rho)$ terms in
the correlations between $\vec X_1$ and $\vec X_2$ (see
(\ref{eq:X1_2variances})). As a result,
\begin{eqnarray}
\label{eq:aveA2squaredandaveA4}%
E\left[\langle A_2\rangle^2\right] &=& 4N^2(1+M^2) + {\cal O}(\rho)%
\nonumber \\%
E\left[\langle A_4\rangle\right] &=& \langle E\left[
A_4\right]\rangle %
\\ \nonumber %
&=& 12N(1+MN)\Tr{\langle \vec U \vec U^\dagger \vec U \vec
U^\dagger \rangle} + {\cal O}(\rho) \\ \nonumber%
&=& 12MN(1+MN) + {\cal O}(\rho)
\end{eqnarray}
Collecting all terms from above and expanding them to ${\cal
O}(\rho^4)$ we obtain the mutual information of
(\ref{eq:DUSTM_small_rho}).

\subsection{Isotropic USTM} \label{app:small_rho_forUSTM}

In the case of USTM, we will expand $I_{USTM}$ to order $\rho^3$.
Here, the analog of (\ref{eq:I_small_rho_dustm_1}) is
\begin{eqnarray}
\label{eq:I_small_rho_iustm_1}%
I &=& \alpha E\left[ \Tr{\vec X^\dagger \vec J_M \vec X}\right]\log_2e \\ \nonumber %
& - & E\left[\log_2 \langle\exp\left(\alpha \Tr{\vec X^\dagger
\vec U \vec J_M \vec U^\dagger \vec X}\right) \rangle \right]
\end{eqnarray}
where $\vec U$ is a $T\times T$ unitary matrix and the $T\times N$
Gaussian random matrix $\vec X$ has the following correlations,
which follow from (\ref{eq:iustm_XX_stat})
\begin{equation}
\label{eq:X_variance_iustm_app}%
E\left[X_{ij}^* X_{kl}\right] = \delta_{ik}\delta_{jl}
(1+J_{ii}\frac{\alpha}{1-\alpha})
\end{equation}
As a result, the first term in (\ref{eq:I_small_rho_iustm_1}) can
be easily evaluated to
\begin{equation}\label{eq:aveXJX}
  \alpha E\left[ \Tr{\vec X^\dagger \vec J_M \vec X}\right] =
  MN\alpha/(1-\alpha) =  TN\rho
\end{equation}
Similarly to the previous section, we define $B_n$ as
\begin{equation}\label{eq:Bn_notation}
  B_n =  \left(\Tr{\vec X^\dagger
\vec U \vec J_M \vec U^\dagger \vec X}\right)^n
\end{equation}
Then, after expanding the second term in
(\ref{eq:I_small_rho_iustm_1})
 to order $\alpha^3$, $I$ becomes
\begin{eqnarray}
\label{eq:I_small_rho_iustm_2}%
I &=& \log_2e \left( MN\frac{\alpha}{1-\alpha} - \alpha
E\left[\langle
B_1\rangle\right]\right.%
\\ \nonumber %
&-&\left.\frac{\alpha^2}{2} \left(E\left[\langle B_2\rangle\right]
- E\left[\langle B_1\rangle^2\right] \right) \right.
\\ \nonumber &-& \left. %
\frac{\alpha^3}{6} \left( E\left[\langle B_3\rangle\right] + 2
E\left[\langle B_1\rangle^3\right] - %
3 E\left[\langle B_1\rangle\langle B_2\rangle\right]
\right) %
\right) \nonumber %
\end{eqnarray}
Using the orthogonality relation for $U(T)$ unitary matrices
\begin{equation}
\label{eq:U_2moment_T}%
\langle U_{ij} U_{lk}^* \rangle = \frac{1}{T}
\delta_{il}\delta_{jk}
\end{equation}
we can calculate $\langle B_1\rangle$ to be
\begin{equation}\label{eq:B1_Uave}
  <B_1> = \frac{M}{T} \Tr{\vec X^\dagger \vec X}
\end{equation}
We can now calculate the terms in (\ref{eq:I_small_rho_iustm_2})
explicitly:
\begin{equation}
\label{eq:small_rho_iustm_mean}%
E\left[\langle B_1\rangle\right] =
MN\left(1+\frac{M}{T}\frac{\alpha}{1-\alpha}\right)
\end{equation}
\begin{eqnarray}
\label{eq:small_rho_iustm_var}%
E\left[\langle B_2\rangle\right] - E\left[\langle
B_1\rangle^2\right] = \langle E\left[B_2\right]\rangle -
E\left[\langle B_1\rangle^2\right]
\\ \nonumber%
 = MN\left(1-\frac{M}{T}\right)\left(1+2\rho\frac{M}{T}\right) +
 {\cal O}(\rho^2)
\end{eqnarray}
\begin{eqnarray}
\label{eq:small_rho_iustm_skew}%
E\left[\langle B_3\rangle\right] + 2
E\left[\langle B_1\rangle^3\right] - %
3 E\left[\langle B_1\rangle\langle B_2\rangle\right]
\\ \nonumber%
 = 2MN\left(1-\frac{M}{T}\right)\left(1-2\rho\frac{M}{T}\right) +
 {\cal O}(\rho)
\end{eqnarray}
Note that the last two equations were only calculated to ${\cal
O}(\rho)$ and ${\cal O}(1)$, given that their proportionality
constants in (\ref{eq:I_small_rho_iustm_2}) are ${\cal O}(\rho^2)$
and ${\cal O}(\rho^3)$, respectively. Collecting all terms
(\ref{eq:small_rho_iustm_mean}), (\ref{eq:small_rho_iustm_var})
and (\ref{eq:small_rho_iustm_skew}) together in
(\ref{eq:I_small_rho_iustm_2}), we get the mutual information for
USTM to ${\cal O}(\rho^3)$ expressed in (\ref{eq:USTM_small_rho}).

\section{Large $\rho$ Analysis}
\label{app:large_rho}

\subsection{Differential USTM}
\label{app:large_rho_forDUSTM}

We wish to calculate the asymptotic behavior of the DUSTM mutual
information for large $\rho$. Using (\ref{eq:aveX1X2}), we rewrite
the log-ratio of (\ref{eq:LLR_DUSTM}) as
\begin{eqnarray}\label{eq:LLR_app_large_rho1}
 && \log_2\left(\frac{p(\vec X|\vec \Phi_0)}{p(\vec X)}\right) =
 \left(2\beta\rho MN - 2\sum_{i=1}^R\sqrt{y_i}\right)\log_2e  \nonumber \\
 &-& \sum_{k=M-R}^{M-1} \log_2 k!  \\ \nonumber %
 &+&\log_2\left[\frac{\det\left(y_j^{M-R+i-1}\right)}{\det\left(y_j^{\frac{M-R+i-1}{2}}
 {\tilde I}_{M-R+i-1}(2\sqrt{y_j})\right)}\right]
\end{eqnarray}
where $y_i$, for $i=1,\cdots,R$ are the $R$ eigenvalues of the
matrix $\beta^2 {\bf  X_2 X_1^\dagger  X_1 X_2^\dagger}$. The
above equation is averaged over the $M\times N$ Gaussian matrices
${\bf X_1}$, ${\bf X_2}$ with correlations given by
(\ref{eq:X1_2variances}). To analyze the large $\rho$ behavior, it
is convenient to use the independent $M\times N$ matrices $\vec
Z_\pm$ with ${\cal CN}(0,1)$ entries, defined as
\begin{eqnarray}
\label{eq:Z_plus_minus_def}%
\vec Z_+ &=& \frac{\vec X_1 +\vec
X_2}{\sqrt{2(1+2\rho)}} \\ \nonumber %
\vec Z_- &=& \frac{\vec X_1 -\vec X_2}{\sqrt{2}}
\end{eqnarray}
Thus $\beta^2 {\bf  X_2 X_1^\dagger X_1 X_2^\dagger}$ can be
written as a sum of terms with decreasing powers of $\rho$:
\begin{equation}\label{eq:expansion_H_n}
\beta^2 {\bf X_1 X_2^\dagger X_2 X_1^\dagger} = \beta^2\rho^2
\left({\bm {\cal H}_0}^2 + \frac{{\bm {\cal H}_1}}{\sqrt{\rho}} +
\frac{{\bm {\cal H}_2}}{\rho}  + {\cal
O}\left(\frac{1}{\rho^{3/2}} \right)\right)
\end{equation}
where
\begin{eqnarray}
\label{eq:H0}%
{\bm {\cal H}}_0 &=& {\bm {\cal N}}_+ \\
\label{eq:H1}%
{\bm {\cal H}}_1 &=& \frac{1}{\sqrt{2}} \left(\left(\vec Z_- \vec
Z_+^\dagger -
\vec Z_+ \vec Z_-^\dagger\right) {\bm {\cal N}}_+ +h.c.\right) \\
\label{eq:H2}%
{\bm {\cal H}}_2 &=& {\bm {\cal N}}_+^2 -%
 \frac{1}{2}\left( {\bm {\cal N}}_+ {\bm {\cal N}}_- + {\bm {\cal
 N}}_- {\bm {\cal N}}_+ \right) \\ \nonumber
 &-& \frac{1}{2}\left(\vec Z_- \vec Z_+^\dagger - \vec Z_+ \vec
 Z_-^\dagger\right)^2
\end{eqnarray}
and ${\bm {\cal N}}_\pm = \vec Z_\pm \vec Z_\pm^\dagger$.

To leading order in $\rho$, we can neglect the higher order terms
in (\ref{eq:expansion_H_n}) and only keep the term proportional to
${\bm{\cal H}}_0^2$. In this case, the eigenvalues of the left
hand side of (\ref{eq:expansion_H_n}) are
$y_i=(\beta\rho\lambda_i)^2$, where $\lambda_i$ are the
eigenvalues of ${\bm {\cal N}}_+$. We will need to calculate $y_i$
to next to leading order, focusing on the $R$ non-zero ones. To do
this we need to express the full eigenvalues $y_i$ as well as
their corresponding eigenvectors as a Taylor expansion in the
small parameter $1/\sqrt{\rho}$. Applying the normalization
condition of the eigenvectors at every order we obtain an
expression for the corrections of the eigenvalues in terms of the
eigenvalues and eigenvectors of the unperturbed matrix, i.e.
${\bm{\cal H}}_0^2$. The perturbation analysis of eigenvalues is
treated in detail in standard textbooks, see for example
\cite{CohenTannoudji_book}. Below we simply quote the answer:
\begin{eqnarray}\label{eq:pert_expansion_yi^2}
y_i &=& \beta^2\rho^2\left(\lambda_i^2 + \frac{\vec n^\dagger_i
{\bm {\cal H}}_1 \vec n_i}{\sqrt{\rho}} + \frac{\vec n^\dagger_i
{\bm
{\cal H}}_2 \vec n_i}{\rho}\right. \\ \nonumber %
 &+& \left.\frac{1}{\rho} \sum_{j\neq i} \frac{\left|\vec n^\dagger_i
{\bm {\cal H}}_1 \vec n_j \right|^2}{\lambda_i^2-\lambda_j^2} +
{\cal O}(\rho^{-3/2}) \right)
\end{eqnarray}
where $\vec n_i$ are the eigenvectors corresponding to
$\lambda_i$. The last term in the above equation is summed over
all $\lambda_j$, including zeros, and is well behaved because the
eigenvalues $\lambda_i$ are unequal with probability $1$. We next
observe that, since $\vec n_i$ are eigenvectors of ${\bm {\cal
H}}_0$, $\vec n^\dagger_i {\bm {\cal H}}_1 \vec n_i=0$.  We now
can expand the second term in (\ref{eq:LLR_app_large_rho1}):
\begin{eqnarray}\label{eq:sqrt_yi_expansion}
  2\sum_{i=1}^R \sqrt{y_i} &=& 2\beta\rho\sum_i^R\left[\lambda_i + \frac{\vec n^\dagger_i {\bm
{\cal H}}_2 \vec n_i}{2\rho\lambda_i} \right. \\ \nonumber
 &+& \left. \frac{1}{2\rho\lambda_i} \sum_{j\neq
i} \frac{\left|\vec n^\dagger_i {\bm {\cal H}}_1 \vec n_j
\right|^2}{\lambda_i^2-\lambda_j^2}+{\cal O}(\rho^{-2}) \right]
\end{eqnarray}
To proceed further, we integrate out $\vec Z_-$ in the above
equation (but not $\vec Z_+$). As a result we get
\begin{eqnarray}\label{eq:sqrt_yi_expansion2}
  2\sum_{i=1}^R \sqrt{y_i} &=& 2\beta\rho\sum_i^R\lambda_i  \\ \nonumber
  &+&   \beta\left(\sum_i^R \lambda_i +(M-N)R\right) + {\cal
   O}(\rho^{-1})
\end{eqnarray}
which, after integrating over $\vec Z_+$ gives
\begin{eqnarray}\label{eq:sqrt_yi_expansion3}
  2E\left[\sum_{i=1}^R \sqrt{y_i}\right] &=& 2\beta\rho MN  \\ \nonumber
  &+&  R\left(M-\frac{R}{2}\right)   + {\cal   O}(\rho^{-1})
\end{eqnarray}
Thus, the first term in the above equation, cancels the first
${\cal O}(\rho)$ term in (\ref{eq:LLR_app_large_rho1}), with the
remainder being only of order unity.

We now turn to the asymptotic treatment of the determinants in
(\ref{eq:LLR_app_large_rho1}). Since for large $\rho$ the non-zero
$y_i$'s will be large, we may use the asymptotic form of the
normalized modified Bessel function
\begin{eqnarray}\label{eq:asymptotic_Bessel_In}
{\tilde I}_n(x) &\equiv& e^{-x} I_n(x) \\ \nonumber
 &\approx& \frac{1}{\sqrt{2\pi x}} (1+ {\cal O}(x^{-1}))
\end{eqnarray}
in the determinant of the denominator in
(\ref{eq:LLR_app_large_rho1}) to obtain
\begin{eqnarray}\label{eq:det_yI(y)_asymptotic}
&& \det\left(
 \frac{y_j^{\frac{M-R+i-1}{2}}}{\sqrt{4\pi\sqrt{y_j}}}(1+{\cal
 O}(y_j^{-1/2}))\right) \nonumber \\
&=& \det\left(
 \frac{\left(\beta\rho\lambda_j\right)^{M-R+i-1}}{\sqrt{4\pi\beta\rho\lambda_j}}(1+{\cal
 O}(\rho^{-1}))\right) \nonumber \\
  &=&
  \frac{\left(\beta\rho\right)^{(2M-R-2)R/2}}{\left(4\pi\right)^{R/2}}
 \prod_{i,j<i}\lambda_i^{M-R-1/2}(\lambda_j-\lambda_i)  \nonumber \\
 &\times& \left[1+{\cal O}(\rho^{-1})\right]
\end{eqnarray}
The first equality follows from the fact that $y_i =
\beta^2\rho^2(\lambda_i^2 + {\cal O}(\rho^{-1})$. Similarly, the
Vandermonde determinant can be expressed as
\begin{eqnarray}
\label{eq:Vandermonde_expansion_large_rho} %
\det(y_j^{M-R+i-1}) &=&
\det\left(\left[\beta^2\rho^2\lambda_j^2(1+{\cal
O}(\rho^{-1}))\right]^{M-R+i-1}\right)  \nonumber \\%
&=& \left(\beta\rho\right)^{R(2M-R-1)} %
 \prod_{i,j<i}\lambda_i^{2(M-R)}(\lambda_j^2-\lambda_i^2)  \nonumber \\
 &\times& \left[1+{\cal O}(\rho^{-1})\right]
\end{eqnarray}
Taking the logarithm of the ratio of the two determinants
(\ref{eq:det_yI(y)_asymptotic}),
(\ref{eq:Vandermonde_expansion_large_rho}), we get
\begin{eqnarray}\label{eq:asymptotic_ratio_of_dets}
\log_2\frac{\det(\cdots)}{\det(\cdots)} &=&
R\left(M-\frac{R}{2}\right)\log\frac{\rho}{2} + \frac{R}{2} \log_2 4\pi \\
\nonumber %
&+& \left(M-R+\frac{1}{2}\right)\sum_{i=1}^R \log_2\lambda_i  \\
\nonumber
 &+& \sum_{i,j<i}\log_2(\lambda_i+\lambda_j) + \log_2(1+{\cal O}(\rho^{-1}))
\end{eqnarray}
Since the eigenvalues of ${\bm {\cal H}}_0$ are equivalent, we
need only to evaluate the averages $E\left[\log_2\lambda_1\right]$
and $E\left[\log_2(\lambda_1+\lambda_2)\right]$ over the $M\times
N$ Gaussian matrix $\vec Z_+$. Careful analysis of the correction
term shows that it is ${\cal O}(\log_2\rho/\rho)$.

To calculate these quantities we need the single eigenvalue
probability density $\rho(\lambda)$ as well as the joint two
eigenvalue probability density $\rho(\lambda_1,\lambda_2)$ for the
random matrix ${\bm {\cal H}}_0 = \vec Z_+^\dagger \vec Z_+$.
Using Telatar's analysis \cite{Telatar1995_BLAST1}, it can be
shown that
\begin{eqnarray}\label{eq:rho_lambda}
\rho(\lambda) &=& \frac{\lambda^Q e^{-\lambda}}{R} \mu_2(\lambda,\lambda) \\
\rho(\lambda_1, \lambda_2) &=&
\frac{\lambda_1^Q\lambda_2^Qe^{-(\lambda_1+\lambda_2)}}{R(R-1)}%
\\ \nonumber
&\times&
\left(\mu_2(\lambda_1,\lambda_1)\mu_2(\lambda_2,\lambda_2)-\mu_2(\lambda_1,\lambda_2)^2\right)
\label{eq:rho_lambda1_2}
\end{eqnarray}
where $\mu_2(\lambda_1, \lambda_2)$ is given by
\begin{equation}\label{eq:mu_2_def}
  \mu_2(\lambda_1,\lambda_2) = \sum_{k=0}^{R-1} \frac{k!}{(k+Q)!}
  L_k^Q(\lambda_1) L_k^Q(\lambda_2)
\end{equation}
and $L_k^Q(x)$ is the associated Laguerre polynomial of order $k$.
Since both $\rho(\lambda)$ and $\rho(\lambda_1,\lambda_2)$ are
finite polynomials in $\lambda_1$, $\lambda_2$ times a exponential
factor, they can be explicitly integrated using the following
identities several times:
\begin{eqnarray}\label{eq:int_log_lambda}
\int_0^\infty d\lambda\, \lambda^n\log_2\lambda e^{-\lambda} &=&n!
\Psi(n+1) \\ \nonumber %
&\equiv& n!\left(1+\frac{1}{2} +\cdots \frac{1}{n} - {\cal
C}\right)\log_2e
\end{eqnarray}
\begin{eqnarray}
\label{eq:int_log_lambda12} %
\int_0^\infty d\lambda_1\, \int_0^\infty d\lambda_2\, \lambda_1^n
\lambda_1^m e^{-(\lambda_1+\lambda_2)}
\log_2(\lambda_1+\lambda_2)&&
\\ \nonumber
= n! m! \Psi(n+m+2)\log_2e &&
\end{eqnarray}
where ${\cal C}$ is the Euler constant ${\cal C}= 0.57721\cdots$.
To somewhat simplify the procedure, we apply the
Christofel-Darboux identity (see \cite{Gradshteyn_Ryzhik_book})
\begin{eqnarray}
\label{eq:Chris_Darboux_12} %
\mu_2(\lambda_1,\lambda_2) &=& \sum_{k=0}^{R-1} \frac{k!}{(k+Q)!}
  L_k^Q(\lambda_1) L_k^Q(\lambda_2) \\ \nonumber
&=& \frac{R!}{(R+Q-1)!} \\ \nonumber %
&\times& \frac{L_{R-1}^Q(\lambda_1) L_R^Q(\lambda_2) -
L_R^Q(\lambda_2) L_{R-1}^Q(\lambda_1)}{\lambda_1-\lambda_2}
\end{eqnarray}
which, in the limit $\lambda_2\rightarrow \lambda_1$ becomes
\begin{eqnarray}\label{eq:Chris_Darboux_11} %
\mu_2(\lambda_1,\lambda_1) &=& \frac{R!}{(R+Q-1)!} \\ \nonumber
&\times& \left(L_{R-1}^Q(\lambda_1) L_{R-1}^{Q+1}(\lambda_1) -
L_R^Q(\lambda_1) L_{R-2}^{Q+1}(\lambda_1)\right)
\end{eqnarray}
Combining (\ref{eq:rho_lambda}), (\ref{eq:int_log_lambda}) and
(\ref{eq:Chris_Darboux_11}), we get
\begin{eqnarray}
\label{eq:ave_log_lambda} %
{\cal L}_1(M,N) = E\left[\log\lambda\right] = \frac{(R-1)!}{(R+Q-1)!} %
&& \\ \nonumber
\times \left[ %
\sum_{k,m=0}^{R-1} (-1)^{k+m} %
\left(\begin{array}{c} Q+R-1\\R-1-k\end{array}\right) %
\left(\begin{array}{c} Q+R\\R-1-m\end{array}\right) %
\right. &&%
\\ \nonumber %
\left. - \sum_{k,m=0}^{R,R-2} (-1)^{k+m} %
\left(\begin{array}{c} Q+R\\R-k\end{array}\right) %
\left(\begin{array}{c} Q+R-1\\R-2-m\end{array}\right)\right]&& %
\\ \nonumber
\times \frac{(Q+k+m)!\Psi(Q+k+m+1)}{k!m!} \log_2e &&
\end{eqnarray}
In the above equation the last term, which appears outside the
bracket, refers to both double sum-terms inside the bracket.
Similarly, by combining (\ref{eq:rho_lambda1_2}),
(\ref{eq:int_log_lambda12}) and (\ref{eq:Chris_Darboux_12}), we
get
\begin{eqnarray}
{\cal L}_2(M,N) = E\left[\log_2\left(\lambda_1+\lambda_2\right)\right] =&& %
\nonumber %
\\ \nonumber %
\left\{\left[\sum_{k_1,m_1=0}^{R-1} (-1)^{k_1+m_1}
\left(\begin{array}{c} Q+R-1\\R-1-k_1\end{array}\right) %
\left(\begin{array}{c} Q+R\\R-1-m_1\end{array}\right) %
\right.\right.&& \\ \nonumber %
-\left.\sum_{k_1 ,m_1=0}^{R,R-2} (-1)^{k_1+m_1}
\left(\begin{array}{c} Q+R\\R-k_1\end{array}\right) %
\left(\begin{array}{c} Q+R-1\\R-2-m_1\end{array}\right) %
\right] && \\ \nonumber %
\times%
\left[\sum_{k_2,m_2=0}^{R-1} (-1)^{k_2+m_2}
\left(\begin{array}{c} Q+R-1\\R-1-k_2\end{array}\right) %
\left(\begin{array}{c} Q+R\\R-1-m_2\end{array}\right) %
\right.&& \\ \nonumber %
-\left.\left.\sum_{k_2,m_2=0}^{R,R-2} (-1)^{k_2+m_2}
\left(\begin{array}{c} Q+R\\R-k_2\end{array}\right) %
\left(\begin{array}{c} Q+R-1\\R-2-m_2\end{array}\right) %
\right]\right\} && \\ \nonumber %
\times \frac{(Q+k_1+m_1)!(Q+k_2+m_2)!}{k_1!k_2!m_1!m_2!}
&& \\ \nonumber %
\times\Psi(2Q+k_1+m_1+k_2+m_2+2)  \log_2e
&& \\ \nonumber %
-\left\{\left[\sum_{k_1,m_1=0}^{R-1,R} (-1)^{k_1+m_1}
\left(\begin{array}{c} Q+R-1\\R-1-k_1\end{array}\right) %
\left(\begin{array}{c} Q+R\\R-m_1\end{array}\right) %
\right. \right.&& \\ \nonumber
\left.\times\frac{sgn(k_1-m_1)}{k_1!m_1!}\sum_{p_1=0}^{|k_1-m_1|-1}
\right] && \\ \nonumber %
\times\left[\sum_{k_2,m_2=0}^{R-1,R} (-1)^{k_2+m_2}
\left(\begin{array}{c} Q+R-1\\R-1-k_2\end{array}\right) %
\left(\begin{array}{c} Q+R\\R-m_2\end{array}\right) %
\right. && \\ \nonumber
\left.\left.\times\frac{sgn(k_2-m_2)}{k_2!m_2!}\sum_{p_2=0}^{|k_2-m_2|-1}
\right]\right\} && \\ \nonumber %
\times (Q+\max(k_1,m_1)-1-p_1+\max(k_2,m_2)-1-p_2)!&& \\ \nonumber
\times(Q+\min(k_1,m_1)+\min(k_2,m_2))!&& \\ 
\times\Psi(2Q+k_1+m_1+k_2+m_2)\log_2e &&
\label{eq:ave_log_lambda12}
\end{eqnarray}
As before, the terms outside the curly brackets are common to all
sums inside the brackets preceding them. After collecting all
terms we can now evaluate the DUSTM mutual information to order
${\cal O}(\log_2\rho/\rho)$.

\subsection{Isotropic USTM}
\label{app:large_rho_forUSTM}

To analyze the large $\rho$ behavior of mutual information of
USTM, we start by writing the mutual information as
\begin{eqnarray}
\label{eq:I_large_rho_iustm_1}%
I_{USTM} &=& E\left[\log_2 \frac{p(\vec X|\vec \Phi_0)}{p(\vec
X)}\right]
\\ \nonumber
&=& \alpha E\left[ \Tr{\vec X^\dagger \vec J_M \vec X}\right] \log_2e\\ \nonumber %
& - & E\left[\log_2 \langle\exp\left(\alpha \Tr{\vec X^\dagger
\vec \Phi \vec \Phi^\dagger \vec X}\right) \rangle \right] \\
\nonumber &=& TN\rho\log_2e - E\left[\log_2
\langle\exp\left(\alpha \Tr{\vec X^\dagger \vec \Phi \vec
\Phi^\dagger \vec X}\right) \rangle \right]
\end{eqnarray}
where the third equality is obtained by integrating over $\vec X$,
see (\ref{eq:aveXJX}). To evaluate the second term we will perform
an asymptotic analysis of the multiple integration in
(\ref{eq:multi_int_iustm_def}), which is performed by evaluating
the residues of the poles of the $t$-integrals. We will assume
that $T>M$, since otherwise the mutual information is identically
zero. We also use the fact that at large $\rho$ from
(\ref{eq:iustm_XX_stat}), the eigenvalues of $\vec X^\dagger \vec
X$ generally split into three groups: the first $R$ being large
${\cal O}(\rho^{-1})$, $K-R$ eigenvalues being ${\cal O}(1)$,
while the remaining $N-K$ being zero. For simplicity, we assume
they are ordered in magnitude, i.e. $y_1\geq y_2\geq\ldots$. Note
first that the last term in (\ref{eq:multi_int_iustm_def})
guarantees that no two $t_i$'s are evaluated at the residue of the
same pole with $y_n\neq 0$. As a result the leading term will
entail $\min(K,M)$ $t$'s evaluated at the poles of the ${\cal
O}(\rho)$ eigenvalues of $\vec X^\dagger \vec X$. All other terms
will be exponentially smaller. Let us start with the simpler case
of $M<K$. Here the $M$ $t$-integrals are all performed by taking
their residues at the $M$ ${\cal O}(\rho)$ $y_i$'s. Thus we get
\begin{eqnarray}
\label{eq:multi_int_iustm_1}%
&&\langle \exp\left(\alpha \Tr{{\bf X^\dagger \Phi \Phi^\dagger
X}}
\right) \rangle  \\ \nonumber %
&\approx&C_{TM} \prod_{m=1}^M \frac{e^{\alpha
y_m}}{\prod_{q=1,q\neq m}^K (\alpha
y_m -\alpha y_q)(\alpha y_m)^{T-K}} \\ \nonumber %
&& \times \prod_{l<m}(\alpha y_l-\alpha y_m)^2\\ \nonumber %
&=& C_{TM} \prod_{m=1}^M\left[\frac{e^{\alpha
y_m}}{\prod_{q=m+1}^K (\alpha
y_m -\alpha y_q)(\alpha y_m)^{T-K}}\right] \\ \nonumber %
&\approx& C_{TM} \prod_{m=1}^M\left[\frac{e^{\alpha y_m}}{(\alpha
y_m)^{T-M}}\right] \left(1+{\cal O}(\rho^{-1})\right)
\end{eqnarray}
where in the last step we used the fact that the eigenvalues
$y_{M+1},\ldots,y_K$ are ${\cal O}(1)$, while $y_m$ for
$m=1,\dots,M$ are $y_m={\cal O}(\rho)$. Thus for $M\leq K$ the
mutual information can be written as
\begin{eqnarray}
\label{eq:I_large_rho_iustm_2}%
I_{USTM} = \left[TN\rho - \alpha \sum_{m=1}^M E\left[
y_m\right]\right]\log_2e - \log_2
C_{T,M} \\ \nonumber %
+ (T-M)\sum_{m=1}^M E\left[\log_2(\alpha y_m)\right] + {\cal
O}(\log_2\rho/\rho)
\end{eqnarray}
Using a similar analysis as in Appendix
\ref{app:large_rho_forDUSTM}, it can be shown that for $M\leq K$
\begin{equation}\label{eq:ave_y_m_large_rho_iustm}%
\alpha \sum_{m=1}^M E\left[ y_m\right]  = TN\rho +(T-M)R+ {\cal
O}(\rho^{-1})
\end{equation}
To calculate the expectation of $\log_2 \alpha y_m$, we note that
to leading order we have $y_m\approx \rho T \lambda_m/M +{\cal
O}(1)$, where $\lambda_m$ are the eigenvalues of $\vec X^\dagger_1
\vec X_1$, with $\vec X_1$ a $M\times N$ Gaussian random,
unit-variance matrix. Thus we can use (\ref{eq:ave_log_lambda}).

When $M>K\equiv N$, we have the added complexity that only $K$
$y_n$'s are ${\cal O}(\rho)$. After performing the first K
$t$-integrals by evaluating them at the poles of these $K$ ${\cal
O}(\rho)$ $y$'s, (\ref{eq:multi_int_iustm_def}) becomes
\begin{eqnarray}
\label{eq:multi_int_iustm_2}%
&&\langle \exp\left(\alpha \Tr{{\bf X^\dagger \Phi \Phi^\dagger
X}}
\right) \rangle  \\ \nonumber %
&\approx&\frac{C_{TM}}{(M-K)!} \prod_{m=1}^K \frac{e^{\alpha
y_m}}{(\alpha y_m)^{T-K}} \\ \nonumber %
&& \times \prod_{m=K+1}^M
\int\frac{dt_m}{2\pi}\frac{e^{-i\lambda_m} \prod_{q=1}^K (-\alpha y_q-i t_m)}{(-i\lambda_m)^{T-K}} \\ \nonumber %
&& \times \prod_{l>m}(-i\lambda_m-i\lambda_l)^2\\ \nonumber %
\end{eqnarray}
The $M-K$ remaining integrals have high-order poles at zero. It is
straightforward to show that the above equation becomes
\begin{eqnarray}
\label{eq:multi_int_iustm_3}%
&&\langle \exp\left(\alpha \Tr{{\bf X^\dagger \Phi \Phi^\dagger
X}}\right) \rangle  \\ \nonumber %
&\approx& C_{TM} \prod_{m=1}^K \frac{e^{\alpha y_m}}{(\alpha
y_m)^{T-M}}\left|\det \vec G\right| \left(1+{\cal
O}(\rho^{-1})\right)
\end{eqnarray}
where $\vec G$ is an ($M-K$)-dimensional square Hankel matrix with
elements
\begin{eqnarray}
\label{eq:G_matrix_large_rho_iusmt}%
G_{mn} = \left\{ \begin{array}{cc} %
\frac{1}{(T-K-m-n+1)!} & m+n\leq T-K+1 \\
0 & \mbox{otherwise} %
\end{array} \right.
\end{eqnarray}
As a result, for $M>N$ and large $\rho$ the mutual information is
asymptotically equal to
\begin{eqnarray}
\label{eq:I_large_rho_iustm_3}%
I_{USTM} &=& (T-M)R\left[\log_2\frac{\rho T}{Me} +{\cal
L}_1\right]- \log_2
C_{T,M} \nonumber \\  %
&-& \log_2\left|\det\vec G\right| + {\cal O}(\log_2\rho/\rho)
\end{eqnarray}

\end{document}